\documentclass{aa}  
\usepackage[varg]{txfonts}
\usepackage{multirow}
\usepackage{graphicx}
\usepackage{color}
\usepackage[colorlinks, citecolor={blue}]{hyperref}

\begin{document}

\title{SIT~45: An interacting, compact, and star-forming isolated galaxy triplet.} 
\titlerunning{SIT~45: An interacting, compact, and star-forming isolated galaxy triplet.}

\author{
D. Grajales-Medina\inst{1,2}
\and 
M.~Argudo-Fernández\inst{3,4,5}
\and
P.~V\'asquez-Bustos\inst{5}
\and 
S.~Verley\inst{3,4}
\and 
M.~Boquien\inst{6}
\and
S.~Salim\inst{7}
\and
S.~Duarte~Puertas\inst{8,9}
\and 
U.~Lisenfeld\inst{3,4}
 \and
D.~Espada\inst{3,4}
\and
H.~Salas-Olave\inst{6}
}

\institute{
Universidad Internacional de Valencia. Carrer del Pintor Sorolla 21, 46002, Valencia, Spain
\and
Facultad de Ingenier\'ia, Departamento de Ciencias B\'asicas, Universidad Ean, Bogot\'a, Colombia
\and
Departamento de F\'isica Te\'orica y del Cosmos Universidad de Granada, 18071 Granada, Spain \\ \email{margudo@ugr.es}
\and 
Instituto Universitario Carlos I de F\'isica Te\'orica y Computacional, Universidad de Granada, 18071 Granada, Spain
\and
Instituto de F\'isica, Pontificia Universidad Cat\'olica de Valpara\'iso, Casilla 4059, Valpara\'iso, Chile. 
\and
Centro de Astronom\'ia (CITEVA), Universidad de Antofagasta, Avenida Angamos 601 Antofagasta, Chile
\and
Department of Astronomy, Indiana University, Bloomington, IN, 47404
\and
D\'epartement de Physique, de G\'enie Physique et d'Optique, Universit\'e Laval, and Centre de Recherche en Astrophysique du Qu\'ebec (CRAQ), Qu\'ebec, QC, G1V 0A6, Canada
\and
Instituto de Astrof\'{\i}sica de Andaluc\'{\i}a - CSIC, Glorieta de la Astronom\'{\i}a s.n., 18008 Granada, Spain
}

   \date{Received ; accepted }

 
  \abstract
   {The underlying scenario of the formation and evolution of galaxy triplets is still uncertain. Mergers of galaxies in isolated triplets give us the opportunity to study the already complex merging process, with minimal contamination of other environmental effects that potentially allow and accelerate galaxy transitions from active star forming to passive galaxies.}
   {The merging system SIT~45 (UGC~12589) is an unusual isolated galaxy triplet, consisting of three merging late-type galaxies, out of 315 systems in the SIT (\textbf{S}DSS-based catalogue of \textbf{I}solated \textbf{T}riplets). The main aims of this work are to study its dynamical evolution and star formation history (SFH), as well as its dependence on its local and large-scale environment.}
   {To study its dynamics, parameters such as the velocity dispersion ($\sigma_{v}$), the harmonic radius ($R_{H}$), the crossing time ($H_0t_c$), and the virial mass ($M_{vir}$), along with the compactness of the triplet ($S$) were considered. To investigate the possible dependence of these dynamical parameters on the environment, the tidal force $Q$ parameters (both local and large-scale) and the projected local density ($\eta_k$) were used. To constrain the SFH, we used CIGALE to fit its observed spectral energy distribution using multi-wavelength data from the ultraviolet to the infrared.}
   {SIT~45 is one of the most compact triplets in the SIT, and it is also more compact than triplets in other samples. According to its SFH, SIT~45 presents star-formation, where the galaxies also present recent ($\sim $200\,Myr) star-formation increase, indicating that this activity may have been triggered by the interaction. Its dynamical configuration suggests that the system is highly evolved in comparison to the SIT. However this is not expected for systems composed of star-forming late-type galaxies, based on observations in compact groups.}
   {We conclude that SIT~45 is a system of three interacting galaxies that are evolving within the same dark matter halo, where its compact configuration is a consequence of the on-going interaction, rather than due to a long-term evolution (as suggested from its $H_0t_c$ value). We consider two scenarios for the present configuration of the triplet, one where one of the members is a tidal galaxy, and another where this galaxy arrives to the system after the interaction. Both scenarios need further exploration. The isolated triplet SIT~45 is therefore an ideal system to study short timescale mechanisms ($\sim 10^8$\,years), such as starbursts triggered by interactions which are more frequent at higher redshift.}

   \keywords{galaxies: general -- galaxies: formation -- galaxies: evolution -- galaxies: interactions -- galaxies: star formation}

   \maketitle
%

\section{Introduction}
\label{Sec:intro}

It is difficult to overstate the importance of galaxy interactions on the evolution of galaxies across cosmic times. Over the past decades, great progress has been made to understand the mechanisms at play in the already complex case of galaxy pairs \citep{2007A&A...468...61D,2010MNRAS.407.1514E,2012MNRAS.423.1544S,2013MNRAS.430.3128E,2019MNRAS.487.2491E,2020MNRAS.494.4969P,2022A&A...664A.125M}. In general, galaxy mergers are an important mechanism affecting galaxies, and may lead to complex stellar populations, due to multiple starburst episodes \citep{1987AJ.....93.1011K,2000ApJ...530..660B,2003A&A...405...31B,2008MNRAS.388.1537M,2008MNRAS.389.1275D,2012A&A...548A.117Y, 2022MNRAS.tmp.2360R}. During galaxy interactions, gas and stars are carried outside of their parent galaxies due to tidal forces or collisions between gas clouds \citep{1977egsp.conf..401T,1992ARA&A..30..705B,1999ApJ...523L.133N,2006MNRAS.372..839N,2014A&A...566A..97J,2021MNRAS.503.2866D}. Tidal tails and bridges can form out of gas, sometimes in huge clouds, and stars can be found far away from their parent galaxies \citep{2008AJ....135..548D,2013LNP...861..327D,2021ApJ...923L..21P}. In some particular cases leading to the formation
of tidal dwarf galaxies \citep{1994A&A...289...83D,2001AJ....121.2524M,2005ApJ...619L..91N,2006A&A...456..481B,2015A&A...584A.113L}.

Despite their importance, the exploration of the merging process in the even more complex, but also interesting case of galaxy triplets have not been studied in detail, in contrast to close pairs or denser groups of galaxies (e.g., compact groups), where more extensive work has been done \citep{1991ApJS...76..153R,1993ApJ...407..448Z,1994A&A...285..404M,2007AJ....133.2630C,2009ApJ...706...67R,2011MNRAS.418.2043E,2014MNRAS.438.1894T,2014A&A...567A.132W,2015MNRAS.450.2593V}. According to \citet{1999A&A...348..113H}, strongly interacting galaxy triplets may be responsible for the formation of a BL Lac object. \citet{2006A&A...446..447R} found a connection between the physical properties of a galaxy triplet and the formation of a giant spiral polar ring galaxy, another peculiar object, from tidal transfer of mass from the other two member galaxies.  A possible on-going triple galaxy merger was found in Hickson compact group 95 (HCG~95) by \citet{1997ApJ...489L..13I}, nevertheless since this is happening in a group, it can not be strictly compared with a merger in a galaxy triplet.

From the point of view of simulations, adding a third galaxy greatly increases the number of possible scenarios in their formation and evolution, making their study considerably more challenging \citep{1968SvA....11.1006A,2001MNRAS.326.1412A}. However, galaxy triplets would be the simplest system to figure out the behaviour in more complex ones such as compact groups, where gas can be expelled by the collisions to the intergalactic medium and it can interact with other member galaxies \citep{2000ApJ...542L..83G,2002A&A...394..823L,2004A&A...426..471L}. On the contrary, galaxy triplets are generally located in low-density environments, where the in-situ interaction between the members would be the main process driving their evolution \citep{2016MNRAS.459.2539C}. This has been also observer in compact groups, as in the Stephan's Quintet \citep{2019A&A...629A.102D}. In addition, the study of isolated merging triplets gives us the possibility to segregate the merging process itself. That is, to separate it from other environmental effects which could enable and accelerate galaxy transitions from actively star forming to quiescent galaxies, such as ram-pressure stripping of the cold interstellar medium \citep{1972ApJ...176....1G}, the removal of  the hot gas halo, also known as `strangulation' \citep{1980ApJ...237..692L,2000MNRAS.318..703B}, or fast encounters with other galaxies \citep[or galaxy `harassment', see also][]{1972AJ.....77..288G,1996Natur.379..613M,1998ApJ...495..139M}.

Galaxy triplets are not common objects in the local Universe. The expected number of triplet galaxy mergers occurring at the present epoch is very low \citep[$<10\%$ of galaxy triplets,][]{2001MNRAS.326.1412A}, however some effort has been done to identify and characterise the observational and dynamical properties of these unusual objects \citep{1979AISAO..11....3K,2000ARep...44..501K}. The advent of large photo-spectroscopic surveys, such as the Sloan Digital Sky Survey \citep[SDSS;][]{2000AJ....120.1579Y,2019ApJS..240...23A}, has allowed us to better characterise these systems \citep{2009MNRAS.394.1409E,Hern_ndez_Toledo_2011,2012MNRAS.421.1897O,2015A&A...578A.110A}, and therefore better understand their formation and evolution \citep{2013MNRAS.433.3547D,2015MNRAS.447.1399D,2016MNRAS.459.2539C,2019MNRAS.482.2627T}. The largest sample of spectroscopically selected galaxy triplets is composed of about one thousand systems \citep{2012MNRAS.421.1897O}. Nevertheless, isolated triplets are even more rare and amount to 315 in the local Universe \citep{2015A&A...578A.110A}. These isolated triplets (i.e., with no physical neighbour within a projected distance of 1\,Mpc and line-of-sight velocity difference $\Delta v \leq$\,500\,km\,s$^{-1}$) compose the SDSS-based catalogue of Isolated Triplets (hereafter SIT). The SIT was compiled by \cite{2015A&A...578A.110A}, based on the tenth data release of the SDSS \citep[SDSS-DR10,][]{2014ApJS..211...17A} and it represents about 3\% of the total number of galaxies in SDSS spectroscopic sample.

The isolated triplet SIT~45 (z\,=\,0.034, $\sim$145\,Mpc), also known as UGC~12589, is a major `wet' merger, i.e. a merger between similar mass, gas-rich, and blue late-type galaxies, involving three member galaxies. Wet mergers are found to be more prominent in low-density environments \citep{2001PASJ...53..395B,2009MNRAS.400.1264S,Lin_2010}. However, we have only found 9 ongoing mergers in the SIT ($<3\%$ of the SIT), where only 3 of them can be classified as wet mergers, including SIT~45. This number is even slower than expected according to \citet{2001MNRAS.326.1412A}. SIT~45 is therefore an ideal and relatively unique object to study wet mergers in an extremely low density environment. 
The existence of these kind of systems can be considered as a consequence of structure formation, which happens more slowly and at smaller scales than in regions with average density, and a possible pathway for the formation of gas rich disks \citep{10.1093/mnras/stw2841}. 

Given that SIT~45 is a system consisting of three interacting galaxies, it is expected to show a complex dynamics and star formation history (SFH).  We therefore focus this work on the study of the evolution of SIT~45 through its dynamic properties and configuration, as well as depending on its local environment and large-scale structure (LSS). We also investigate its SFH using multi-wavelength data, from the ultraviolet (UV) to the mid-infrared (MIR).
This study is organised as follows. In Sec.~\ref{Sec:data} we describe in detail the isolated triplet SIT~45 and some of its properties in comparison with the SIT (including environment), as well as the dynamical parameters used in this study and the data used to get its spectral energy distribution (SED) to constrain its SFH. We present our results in Sec.~\ref{Sec:res} and the associated discussion in Sec.~\ref{Sec:dis}. Finally, a summary and the main findings of the study are presented in Sec.~\ref{Sec:con}. Throughout the study, a cosmology with $\Omega_{\Lambda_{0}} = 0.7$, $\Omega_{\rm m_{0}} = 0.3$, and $H_0=70$\,km\,s$^{-1}$\,Mpc$^{-1}$ is assumed.


\section{Data and methodology}
\label{Sec:data}

\subsection{The unusual system SIT~45}
\label{Sec:SIT45}

\begin{figure}
\centering
\includegraphics[width=\columnwidth,trim={2cm 2cm 2cm 2cm},clip]{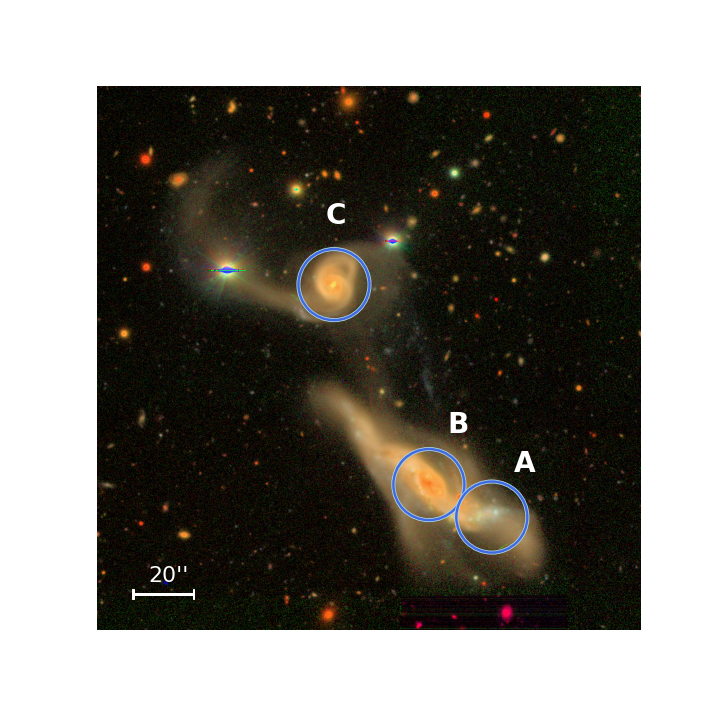}
\caption{Hyper Suprime-Cam \citep[HSC,][]{2022PASJ...74..247A} g-r-i band colour image of the isolated triplet SIT~45 (UGC~12589). North is up and east is left. The A, B, and C galaxies are marked in their central region with a light blue circle and the corresponding letter. The 20\arcsec scale of the image corresponds to 0.4 arcsec per pixel, with a total physical extent of $\sim$170\,kpc.}
\label{fig:SIT45_SDSS_grid}
\end{figure}

According to the NASA/IPAC Extragalagtic Database (NED\footnote{\url{https://ned.ipac.caltech.edu/}}), SIT~45 (UGC~12589) was initially classified as a galaxy pair \citep{1991rc3..book.....D}, composed of SIT~45B and SIT~45C separated by a projected distance of $\sim$52\,kpc. However \citet{2015A&A...578A.110A} noted the existence of SIT~45A, at a projected distance of $\sim$17\,kpc from SIT~45B (as shown in Fig.~\ref{fig:SIT45_SDSS_grid}). The line of sight velocity difference between SIT~45A and SIT~45B is $\Delta\,v~\simeq~118$\,km\,s$^{-1}$, and $\Delta\,v~\simeq~104$\,km\,s$^{-1}$ between SIT~45A and SIT~45C, with a projected separation of $d~\simeq~64$\,kpc. 

Table~\ref{tab:SIT45_general} shows general properties of the galaxies composing SIT~45: coordinates, redshift, apparent size, and absolute magnitudes.

Galaxies in the SIT were selected from a volume—limited sample of galaxies in the SDSS-DR10 in the redshift range $0.005~\leq~z~\leq~0.080$ and apparent magnitude $11~\leq~m_{r}~\leq~15.7$, where $m_{r}$ is the SDSS \textit{model} magnitude in $r$ band. The SIT is composed of 315 physically bound isolated triplets within a projected distance up to $d~\leq~450$\,kpc with a light-of-sight velocity difference $\Delta\,v~\leq~160$~km~s$^{-1}$. More details about the criteria to select physically bound systems can be found in \citet{2015A&A...578A.110A}. Note that galaxies in the SIT are named according to their $m_{r}$ apparent magnitude, with galaxy A (or central galaxy) the brightest and galaxy C the faintest.

\begin{table*}
    \centering
    \begin{tabular}{cccccc}
    \hline \hline
    (1) & (2) & (3) & (4) & (5) & (6) \\
     Galaxy & RA  & DEC  & z & $r_{90}$ & (M$_u$, M$_g$, M$_r$, M$_i$, M$_z$) \\ 
      & (deg) & (deg) &  & (arcsec) & (abs mag) \\
      \hline
     SIT~45A & 23:25:00.25 & -00:00:08.1 & 0.03358 & 28.51 & (-19.37,-20.11,-20.39,-20.45,-20.43) \\
     SIT~45B & 23:25:01.84 & -00:00:01.5 & 0.03397 & 26.03 & (-18.85,-19.62,-20.00,-20.21,-20.04) \\
     SIT~45C & 23:25:03.81 & +00:01:07.2 & 0.03392 & 1.44 & (-16.69,-17.52,-18.18,-18.59,-18.84) \\ 
     \hline
    \end{tabular}
    \caption{General properties of SIT~45. The columns correspond to: (1) name of the galaxy in SIT~45; (2) J2000.0 right ascension in degrees; (3) J2000.0 declination in degrees; (4) redshift of the galaxy; (5) Petrosian radius containing 90\,\% of the total flux of the galaxy in the $r$-band; (6) absolute magnitude in the SDSS-DR10 ($u$, $g$, $r$, $i$, $z$) bands. Galaxy coordinates, redshift, and $r_{90}$ data are from the SDSS-DR10. The absolute magnitudes were obtained by fitting the spectral energy distribution in the five bands of the SDSS-DR10 using the k-correct routine \citep{2007AJ....133..734B}.}
    \label{tab:SIT45_general}
\end{table*}

\subsection{Multiwavelength UV to IR broadband SED fitting}
\label{Sec:SED}

\begin{figure*}
    \centering 
    \includegraphics[width=.4\columnwidth,trim={3cm 1cm 3cm 1cm},clip]{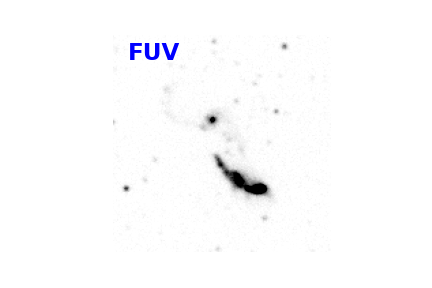}
    \includegraphics[width=.4\columnwidth,trim={3cm 1cm 3cm 1cm},clip]{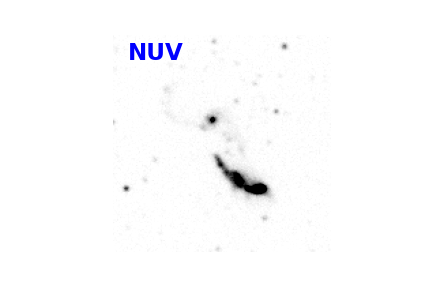} \\
    \includegraphics[width=.4\columnwidth,trim={3cm 1cm 3cm 1cm},clip]{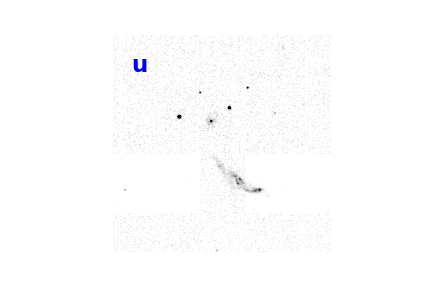}
    \includegraphics[width=.4\columnwidth,trim={3cm 1cm 3cm 1cm},clip]{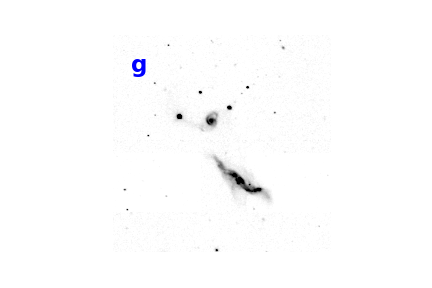}
    \includegraphics[width=.4\columnwidth,trim={3cm 1cm 3cm 1cm},clip]{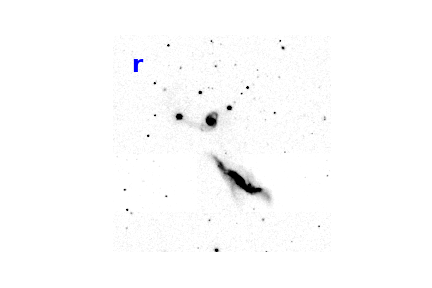}
    \includegraphics[width=.4\columnwidth,trim={3cm 1cm 3cm 1cm},clip]{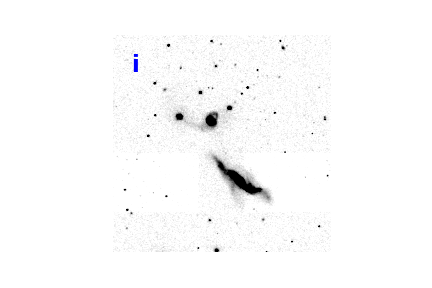}
    \includegraphics[width=.4\columnwidth,trim={3cm 1cm 3cm 1cm},clip]{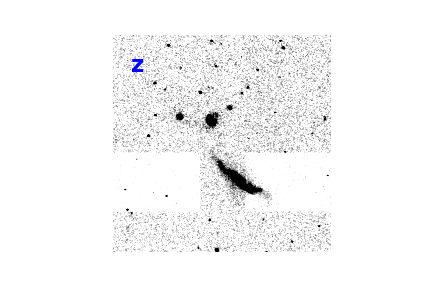} \\
    \includegraphics[width=.4\columnwidth,trim={3cm 1cm 3cm 1cm},clip]{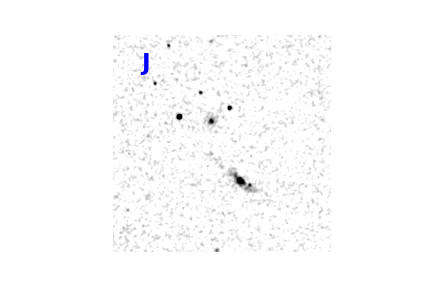}
    \includegraphics[width=.4\columnwidth,trim={3cm 1cm 3cm 1cm},clip]{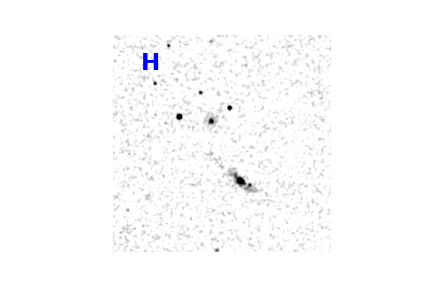}
    \includegraphics[width=.4\columnwidth,trim={3cm 1cm 3cm 1cm},clip]{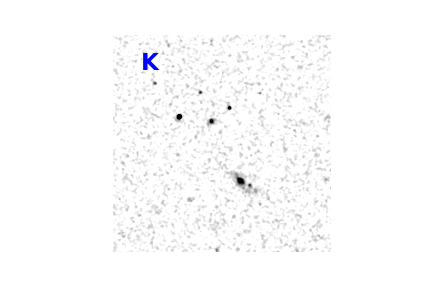} \\
    \includegraphics[width=.4\columnwidth,trim={3cm 1cm 3cm 1cm},clip]{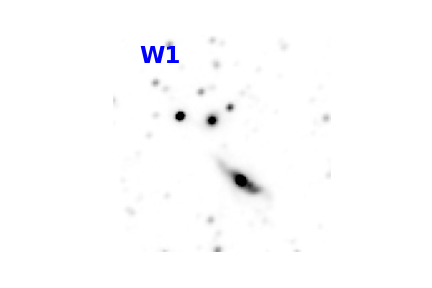} 
    \includegraphics[width=.4\columnwidth,trim={3cm 1cm 3cm 1cm},clip]{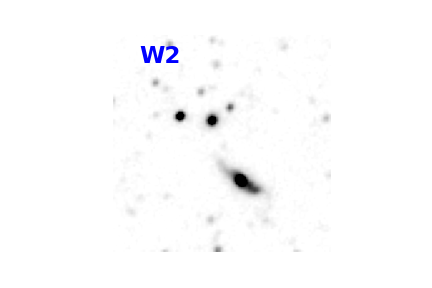}
    \includegraphics[width=.4\columnwidth,trim={3cm 1cm 3cm 1cm},clip]{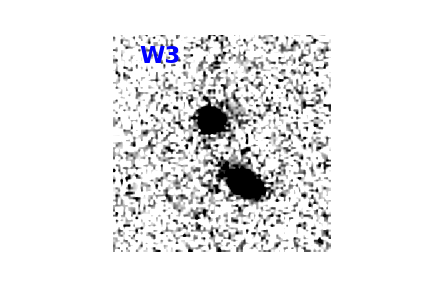}
    \caption{Multiwavelength (UV-optical-NIR) 4\,arcmin$^2$ FoV images of SIT~45. GALEX (FUV and NUV), SDSS (u, g, r, i, and z), 2MASS (J, H, and K), and unWISE (W1, W2, and W3) are presented from left upper to right lower panels, respectively. All the images are rotated north-up and east-left.}
    \label{fig:multilambda}
\end{figure*}

We used multiwavelength photometry, from the ultraviolet (UV) to infrared (IR), to derive the panchromatic spectral energy distribution (SED) for SIT~45. In combination with models, the SED of a galaxy is a powerful tool to constrain key physical properties of the unresolved stellar populations \citep{2011Ap&SS.331....1W}. The UV data were taken from the Galaxy Evolution Explorer (GALEX) satellite All Sky Survey \citep{2005ApJ...619L...1M}. In particular, we used the photometry with the deepest depth of UV observations, both in the far-UV (FUV), at 1528\,$\AA$, and the near-UV (NUV), at 2271\,$\AA$. The optical comes from the SDSS-DR10 photometry in the five SDSS bands: $u$ (3551 $\AA$), $g$ (4686 $\AA$), $r$ (6166 $\AA$), $i$ (7480 $\AA$), and $z$ (8932 $\AA$). The IR photometry comes from the Wide-field Infrared Survey Explorer \citep[WISE;][]{2010AJ....140.1868W}. In particular we used the unofficial unblurred coadds of the WISE imaging \citep[unWISE;][]{2016AJ....151...36L}, which provide significantly deeper imaging while preserving the resolution of the original WISE
images. We used the W1, W2, and W3 WISE bands, centred at 3.4, 4.6, and 12 $\mu$m, respectively. We also used near-infrared (NIR) photometry from the Two Micron All Sky Survey \citep[2MASS;][]{2006AJ....131.1163S}. The cutout images of the galaxies in SIT~45 in each band are shown in Fig.~\ref{fig:multilambda}. In order to combine the photometry in the different bands, we used the high resolution convolution kernels developed by \citet{2011PASP..123.1218A} to convolve to the instrumental point-spread function (PSF) of the W3 band (PSF~$\simeq$~6\arcsec).

We used CIGALE\footnote{\url{https://cigale.lam.fr/}} \citep[Code Investigating GALaxy Emission;][]{2005MNRAS.360.1413B,2009A&A...507.1793N,2019A&A...622A.103B} to perform the UV to IR broadband SED fitting, since it is a code specifically developed to study galaxy emission taking into account both the UV/optical dust attenuation as well as its re-emission in the infrared. Hence, this code has been widely used in the literature to derive star formation rate (SFR), star formation history (SFH) and, dust attenuation in galaxies from the multi-wavelength SED UV to IR \citep{2011A&A...533A..93B, 2011A&A...525A.150G,2014A&A...571A..72B,2017A&A...608A..41C,2018A&A...613A..13Y,2022arXiv220211723B}. A detailed description of the code can be found in \citet{2019A&A...622A.103B}.

Given the nature of SIT~45, we assumed a delayed SFH with optional constant burst/quench as explained in \citet{2017A&A...608A..41C}. They expanded the delayed SFH allowing for an instantaneous recent variation of the SFR, which is flexible to model star-forming galaxies inside the scatter of the main sequence \citep{2007ApJ...660L..43N,2011A&A...533A.119E}, but also starbursts and rapidly
quenched galaxies \citep{2016A&A...585A..43C}. This is therefore one of the most simple methods to model the SFH of interacting galaxies, inasmuch as it considers an older stellar population with an upward or downward recent variation of the SFH, setting it to a constant until the last time step. This module is provided as \texttt{sfhdelayedbq} in CIGALE \citep[see][for details]{2019A&A...622A.103B}.

To compute the analytic SFH, we adopt the single stellar populations of \citet{2003MNRAS.344.1000B}, hereafter BC03, considering a Salpeter IMF \citep{1955ApJ...121..161S}, and a fixed value of the stellar metallicity $\rm Z~=~0.02$ (solar metallicity) to model the non-attenuated stellar emission. The energy absorbed by the dust, considering a modified version of the dust attenuation model of \citet{2000ApJ...533..682C} with a flexible attenuation curve \citep{2009A&A...507.1793N}, is re-emitted in the IR using the dust emission templates of \citet{Dale_2014}, without considering AGN contribution. Note that we have checked that galaxies in SIT~45 are non AGNs using the analysis of \citet{2004MNRAS.351.1151B} on SDSS spectra. The values of the different parameters we used to create a set of SED models are listed in Table~\ref{tab:SEDparams}. These are the parameters used in CIGALE to model the SFH, dust attenuation, and dust emission. With these parameters we modelled 1\,088\,640 different SEDs. Note that with the Bayesian analysis performed by CIGALE, we do not select as the best model the one that minimises the chi squared fitting. We therefore provide mean values, and their uncertainties, weighted by the probability density function of all the models.

\begin{table*}
\centering
\begin{tabular}{ccc}
\hline \hline
Model & Parameter & value\\ 
\hline
SFH  & $\rm Age$ (Myr) & 11000, 12000, 13000 \\ 
 & $\tau_{\rm main}$ (Myr) & 1000, 3000, 5000, 7000, 9000 \\
 & $\rm Age_{bq}$ (Myr) & 20, 50, 100, 300 \\
 & $r_{\rm SFR}$ & 0, 0.25, 0.5, 0.75, 1, 1.5, 2, 5, 10 \\ 
\hline
Dust attenuation & $\rm E(B-V)_{\rm lines}$ (mag) & 0.05, 0.10, 0.15, 0.20, 0.25, 0.30, 0.35, 0.40,\\
 & & 0.45, 0.50, 0.55, 0.60, 0.65, 0.70, 0.75, 0.8 \\
 & $\rm E(B-V)_{\rm factor}$  & 0.25, 0.5, 0.75 \\
  & UV bump wavelength (nm) & 217.5\\
  & UV bump width (nm) & 35.0 \\
  & UV bump amplitude & 0.0, 1.5, 3.0 \\
 & $\Delta \delta$  & -1.0,-0.9,-0.8,-0.7,-0.6,-0.5,-0.4,-0.3,-0.2,-0.1,0,0.1,0.2,0.3 \\ 
\hline
Dust emission & $\alpha$ & 2.0 \\
\hline
\end{tabular}
\caption{Parameters used in CIGALE to model the SFH (upper rows), dust attenuation (middle rows), and dust emission (lower row). Meaning of the parameters: a) SFH: $\rm Age$ --  Age of the main stellar population in the galaxy, in Myr; $\tau_{\rm main}$ -- $e$-folding time of the main stellar population model, in Myr; $\rm Age_{bq}$ -- Age of the burst/quench episode, in Myr; and $r_{\rm sfr}$ -- Ratio of the SFR after/before $\rm Age_{bq}$, values larger than one correspond to an enhancement of the SFR whereas values lower than one will correspond
to a decrease; b) Dust attenuation: $E(B-V)_{\rm lines}$ -- Colour excess of the nebular lines light for both the young and old population; $E(B-V)_{factor}$ -- Reduction factor to apply on $\rm E(B-V)_{\rm lines}$ to compute $E(B-V)_{s}$ the stellar continuum attenuation. Both young and old populations are attenuated by $E(B-V)_{s}$; UV bump wavelength -- Central wavelength of the UV bump in nm; UV bump width -- Width (FWHM) of the UV bump in nm; UV bump amplitude -- Amplitude of the UV bump. For the Milky Way: 3; $\Delta \delta$ -- Slope delta of the power law modifying the attenuation curve. c) Dust emission module: $\alpha$ -- $\alpha$ slope in the \citet{Dale_2014} model.}
    \label{tab:SEDparams}
\end{table*}

\subsection{Quantification of the environment}
\label{Sec:SIT}

\citet{2015A&A...578A.110A} provided two parameters to quantify the effects of the local and the LSS environment on each isolated triplet: the tidal strength parameter ($Q$) and the projected number density $(\eta_{k})$. The combination of these two complementary parameters describes the environment around galaxies \citep{2007A&A...472..121V,2013MNRAS.430..638S,2013A&A...560A...9A,2014A&A...564A..94A}. 

The $Q$ parameter of a galaxy is equivalent to the sum of the individual ratios $\frac{\rm F_{tidal}}{\rm F_{bind}}$ of the external tidal force exerted by each neighbouring galaxy ($\rm F_{tidal}$), with respect to its internal binding force ($\rm F_{bind}$). The $Q$ for a galaxy P is defined by the equation \citep{2007A&A...472..121V,2013A&A...560A...9A}: 

\begin{equation}\label{Eq:Q} 
Q \propto \log\left[\sum_{i}\frac{M_{\star i}}{M_{\star P}}\left(\frac{D_{P}}{d_{i}}\right)^{3}\right] \quad,
\end{equation}   
where $M_{\star i}$ and $d_{i}$ are the stellar mass and the projected physical distance of the considered $i^{th}$ neighbour, $M_{\star P}$ and $D_{P}$ are the stellar mass and apparent diameter of galaxy P, where $D_{P} = 2 \alpha r_{90}$ following the empirical calibration in \citet{2013A&A...560A...9A}, with $\alpha=1.43$ and $r_{90}$ the Petrosian radius containing the 90\% of the total flux of the galaxy. 

To consider the local environment, \citet{2015A&A...578A.110A} quantified the effect of galaxies B and C on galaxy A ($Q_{A,trip}$). As we can observe in the case of SIT~45, the brightest galaxy in the system is not always the most massive. We therefore use the $Q_{trip}$ parameter as defined in V\'asquez-Bustos et al. (in preparation), which provides a more complete information of the tidal strengths in galaxy triplets. It not only considers $Q_{A,trip}$, but also the total tidal strengths exerted on the B ($Q_{B,trip}$) and the C ($Q_{C,trip}$) galaxies as follows:
\begin{equation}\label{Eq:Qtrip}
    Q_{trip} = \frac{Q_{A,trip} + Q_{B,trip} + Q_{C,trip}}{3} \quad.
\end{equation}

For the LSS environment, \citet{2015A&A...578A.110A} considered galaxies within a volume of $\Delta\,v~\leq~500$\,km\,s$^{-1}$ from 1 to 5\,Mpc projected radius, that is neighbouring galaxies at distances larger than the isolation criteria definition of the triplets up to 5\,Mpc, defining a $Q_{\rm LSS}$ parameter.

The projected number density $(\eta_{\rm k,LSS})$ is used to characterise the LSS surrounding each isolated triplet, and is defined as:  
\begin{equation}\label{Eq:eta}
\eta_{\rm k,LSS}\equiv \log\left(\frac{k-1}{Vol(d_{k})}\right)=\log\left(\frac{3(k-1)}{4\pi d_{k}^3}\right) \quad,
\end{equation}
where $d_{k}$ is the projected physical distance to the $k^{th}$ nearest neighbour, with $k$ equal to 5, or less, if there are not enough neighbours in the field out to a projected distance of 5\,Mpc. 

According to \citet{2013A&A...560A...9A, 2014A&A...564A..94A}, the most isolated systems with respect to the LSS environment (i. e. very isolated triplets from any external influence) show low values of $\eta_{\rm k,LSS}$ and $Q$ (either $Q_{\rm trip}$ or $Q_{\rm LSS}$). On the contrary, less isolated systems show higher values of these parameters, since they may be affected by perturbations generated by the existence of neighbours in the LSS. Additionally, a galaxy triplet with an average $\eta_{\rm k,LSS}$ and high $Q_{\rm trip}$ is surely presenting on-going strong interaction/mergers. For average $\eta_{\rm k,LSS}$, it could be also likely identified as a compact triplet if this is presenting a high $Q_{\rm LSS}$. Note that the $Q$ parameter is not a measurement of the compactness of a system, but more an indicator of its degree of interaction or isolation.

When working with isolated galaxies or galaxy groups, \citet{2015A&A...578A.110A,2016A&A...592A..30A} have shown that these parameters can be used to identify if the system is mainly located in a void environment (low values of $Q_{\rm LSS}$), or on the contrary, it is closer to the outskirts of larger structures, such as walls, filaments, or clusters, with higher values of $Q_{\rm LSS}$.

\subsection{Dynamical parameters}
\label{Sec:dynparams}

To characterise the dynamics of SIT~45 we have used a series of parameters that are usually defined to study the dynamics of clusters of galaxies or compact groups, and which can also be applied to small galaxy groups, such as triplets \citep{2005KFNT...21a...3V,2009MNRAS.394.1409E}. The dynamical parameters used in this work are the projected mean harmonic radius, $R_{\rm H}$; the velocity dispersion of the triplet, $\sigma_v$; the dimensionless crossing time, $H_{0}t_{c}$; and the virial mass of the triplet, $M_{\rm vir}$. These parameters are defined as follows. 

The parameter $R_H$ is an estimation of the size of the halo of a galaxy group or cluster. It measures the effective radius of its gravitational potential, independently of the location of the centre of the system. The parameter is defined by the equation \citep{2009MNRAS.400.1317A}:

\begin{equation}\label{eq:RH}
R_{\rm H}=\left(\frac{1}{N}\sum{R_{ij}^{-1}}\right)^{-1}\quad,
\end{equation}
where $R_{ij}$ is the projected distance between the galaxies of the system, in kpc, and $N$ is the number of galaxies ($N = 3$). According this definition, $R_H$ is mainly based on the projected separation between the triplet members, indicating whether they are close to each other (that is, its compactness) or far from each other (that is, if the system is a loose group). Therefore, $R_H$ is independent of the location of the triplet centre, allowing it to reflect the internal structure of the system \citep{2009MNRAS.400.1317A,2015MNRAS.453.2718W}.

Velocity dispersion in galaxy groups and clusters commonly form the basis for dynamical estimates of their mass using the virial theorem \citep{1990AJ....100...32B}. In the present study we used the (line-of-sight) velocity dispersion, $\sigma_{v_{los}}$, defined as \citep{2015MNRAS.447.1399D}:

\begin{equation}\label{eq:sigma}
    \sigma_{v_{los}}^2=\frac{1}{N-1}\sum(v_{r}-\langle{v_{r}}\rangle)^{2}\quad,
\end{equation}
where $v_{r}$ is the line-of-sight velocity, and $N = 3$. Under this definition, the velocity dispersion of the triplet is an estimation of how fast member galaxies move between each other. It is important to note that projection effects might be important. We therefore estimated the three-dimensional velocity dispersion (used in the computation of the dynamical parameters $H_{0}t_{c}$ and $M_{\rm vir}$, as shown below) as $\sigma_{v_{3D}}^2 = 3 \sigma_{v_{los}}^2$. Hereafter we generally use $\sigma_{v}$ to refer to $\sigma_{v_{los}}$.

Following the work of \citet{1992ApJ...399..353H}, for compact groups of galaxies, the dimensionless crossing time parameter, $H_{0}t_{c}$, and the virial mass, $M_{\rm vir}$, are defined by the equations: 

\begin{equation}\label{eq:Tc}
H_{0}t_{c}=H_{0}\,\frac{4{R_{H}}}{\sqrt{3}\pi\sigma_{v}}\quad,
\end{equation}
and
\begin{equation}\label{eq:Mvir}
M_{\rm vir}=\frac{3\pi R_{H}\sigma_v^2}{2G}\quad,
\end{equation}
with $G$ is the gravitational constant. According to \citet{1992ApJ...399..353H}, $H_{0}t_{c}$ is the ratio of the crossing time to the age of the universe, where its reciprocal is related to the maximum number of times a galaxy could have traversed the group since its formation.

In addition to these parameters, we have estimated the compactness of the triplet, $S$, to have a measurement of the spatial configuration of the galaxies in SIT~45, complementary with the parameter $R_H$. The compactness is a physical property that measures the percentage of the total area that is filled with the light of the galaxies composing the triplet. This parameter $S$ is defined as:

\begin{equation}\label{eq:SS}
S=\frac{\sum_{i=1}^3{r_{90}^2}}{r_{m}^2}\quad,
\end{equation}
where $r_{m}$ is the minimum radius of the circle containing the geometrical centres of the three galaxies in triplets \citep{2013MNRAS.433.3547D}. We used these parameters to compare SIT~45 to other triplets in the SIT.


\section{Results}
\label{Sec:res}

\subsection{Star formation history of SIT~45}
\label{Sec:resSFH}

\begin{figure}
\centering
\includegraphics[width=\columnwidth, trim={0cm 0cm 1.4cm 1.2cm},clip]{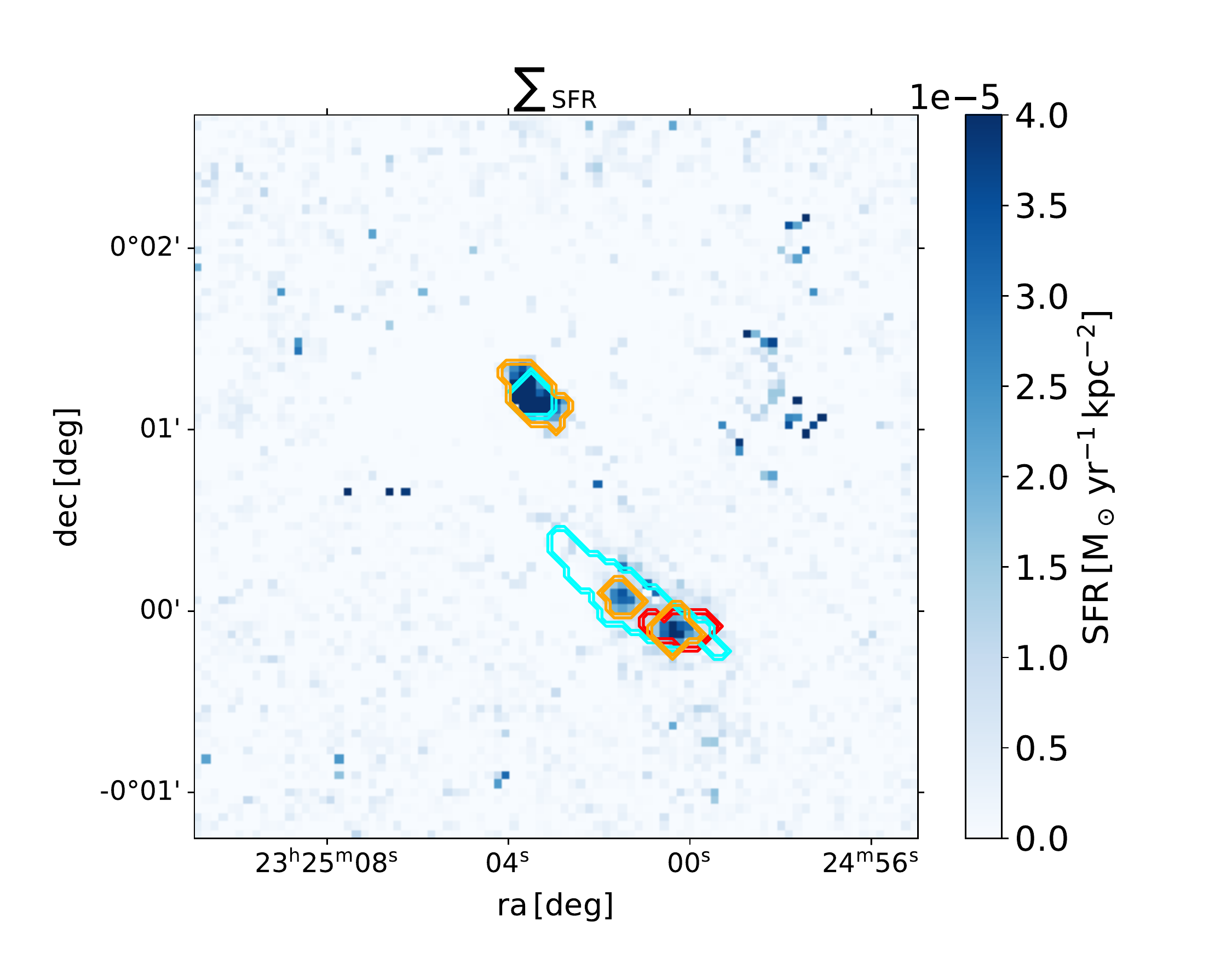}
\caption{SFR surface density ($\rm \sum _{SFR}$) map obtained for the isolated triplet SIT\,45 in a 4\,arcmin$^2$ FoV, North is up, East is left. Coordinates are given in J2000. Colour bar is normalised to show the mean value for all the pixels within 5\,$\sigma$. Orange contours correspond to the mask defined on the $\rm \sum _{SFR}$ map using astrodendro. For an easiest comparison, we show the contours  corresponding to the A galaxy (red contour) and the B and C galaxies (cyan contours). White contours were defined as astrodendro masks on a high resolution three-colour (g,r,i) image from the Hyper Suprime-Cam of the Subaru telescope \citep{2022PASJ...74..247A} of SIT\,45, (as shown in Fig.~\ref{fig:SIT45_SDSS_grid}). The red contour corresponding to the A galaxy selected using the GALEX FUV image (as shown in the upper left panel of Fig.~\ref{fig:multilambda}).}
\label{fig:sfr_map}
\end{figure}

As mentioned in Sec.~\ref{Sec:SED}, we used CIGALE to constrain the SFH of the isolated triplet SIT~45 using the set of values presented in Table~\ref{tab:SEDparams} fitting the observed SED from the UV to the IR.
Since SIT\,45A and SIT\,45B are overlapping in the same area, it is difficult to separate the photometry contribution of each galaxy. We therefore perform the SED fitting to all the pixels within an area of  4\,arcmin$^2$ around the system for each band, obtaining surface densities of the properties of the unresolved stellar populations. In particular, we use the the SFR surface density ($\rm \sum _{SFR}$) map   presented in Fig.~\ref{fig:sfr_map} to select the pixels corresponding to the A, B, and C galaxies.

We use the \textit{astrodendro}\footnote{\url{http://www.dendrograms.org/}} Python package to perform a 'dendrogram' (hierarchical tree-diagram) analysis of the resulted SFR surface density map (as shown in Fig.~\ref{fig:sfr_map}). A dendrogram graphically represents the hierarchical structure of nested isosurfaces, which is particularly useful for analysing astronomical data \citep{2008ApJ...679.1338R,2009Natur.457...63G}. The resulted mask for each galaxy is represented by orange contours. Note the difference with the contours using the same methodology but applied to photometry (represented by cyan and red contours, see image description).
The parameters found for the SFH of each galaxy are shown in Table~\ref{tab:SIT45_bayesian}, following a Bayesian analysis and integrating the pixels corresponding to each galaxy to get total values of the parameters.

\begin{figure}
\centering
\includegraphics[width=\columnwidth,trim={0cm 2cm 1cm 3cm},clip]{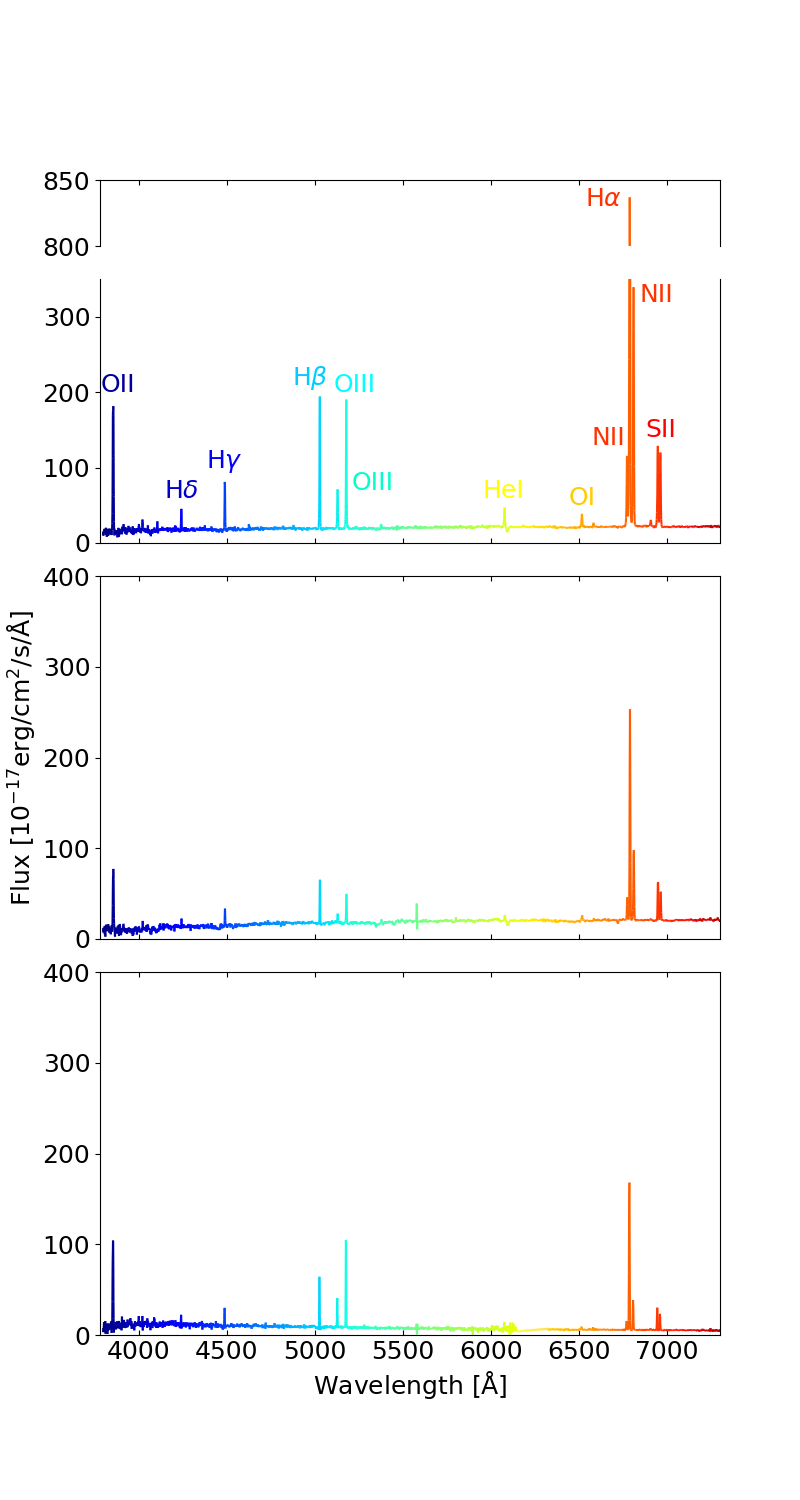}
\caption{SDSS optical spectra of the central region of the galaxies composing SIT~45. Each spectrum is shown for galaxy A, B, and C from upper to lower panel. Spectra are cut at 7500\,$\AA$ to better show the identified emission lines.}
\label{fig:SIT45spectra}
\end{figure}

\begin{figure}
\centering
\includegraphics[width=\columnwidth,trim={0cm 0cm 1cm 0cm},clip]{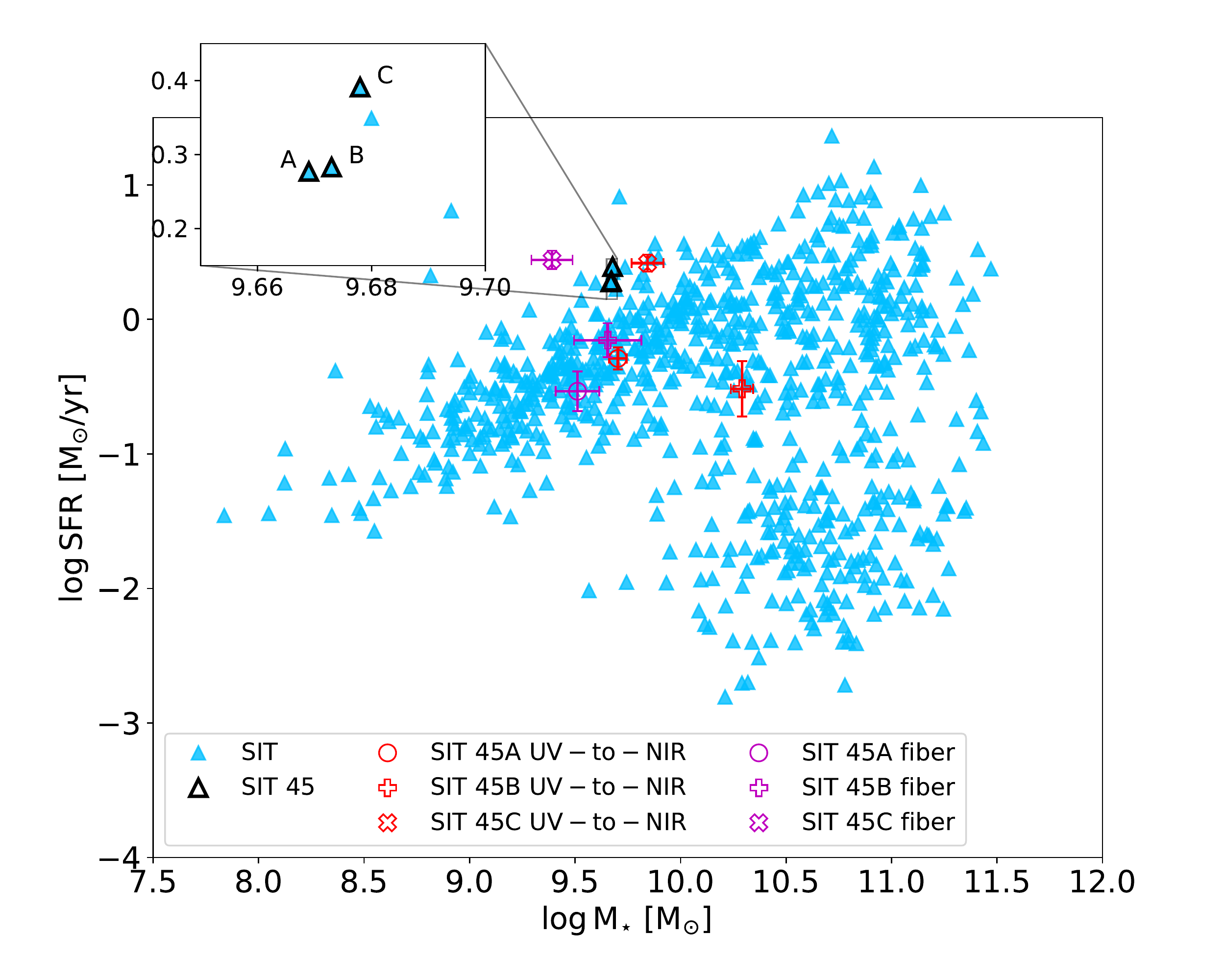}
\caption{Diagram of log(SFR) versus log(M$_{\star}$) for SIT galaxies in the GSWLC-M2 catalogue by \citet{2018ApJ...859...11S} (light blue triangles). The corresponding values for the galaxies in SIT~45 in this catalogue are represented by open black triangles. A zoom is provided to better identify the corresponding values. The values of the SFR and M$_{\star}$ for galaxies SIT~45A, SIT~45B, and SIT~45C using SED fitting from the UV to the IR in this work are represented by red open markers according to the legend, with their corresponding errors using red error bars. We also show the values of the SFR and M$_{\star}$ using SED fitting combining fiber photometry and spectra from the SDSS using magenta open markers according to the legend, with their corresponding errors using magenta error bars.}
\label{fig:SFRmassSIT45}
\end{figure}

In general, the uncertainties found for the parameters are high, suggesting that the parameters of the SFH might be degenerate. However, the results are consistent with the spectroscopic classification of the galaxies in \citet{2004MNRAS.351.1151B} using SDSS spectra. In addition, we also run CIGALE including emission line measurements from the Max Plank Institute for Astrophysics and Johns Hopkins University \citep[MPA-JHU\footnote{Available at \texttt{http://www.mpa-garching.mpg.de/SDSS/DR7/}};][]{2003MNRAS.341...33K,2004ApJ...613..898T,2007ApJS..173..267S} added value catalogue \citep{2004MNRAS.351.1151B} with spectroscopic reanalysis of the optical SDSS spectra of the galaxies (shown in Fig.~\ref{fig:SIT45spectra}). In particular, we used emission line fluxes H$_\alpha$ and H$_\beta$, with the D4000 break parameter using the \citet{1999ApJ...527...54B}, in combination with fiber SDSS magnitudes, using CIGALE. We found that the results for the SFR are consistent among them, as shown in Table~\ref{tab:SFRspec}, with some discrepancies in the stellar mass estimation, as can be observed in Fig.~\ref{fig:SFRmassSIT45}, which is a consequence of using information limited to the optical range and aperture effects.

To compare with the SIT, we use the GALEX-SDSS-WISE Legacy Catalog \citep[GSWLC;][]{2016ApJS..227....2S}, which provides SFRs for 700,000 low-redshift (0.01~<~z~<~0.30) galaxies in the SDSS. The GSWLC also used CIGALE to derive SFR. There are three versions of the catalogue (GSWLC-A, M, D), depending on the depth of the UV photometry. In particular we use the GSWLC-2 catalogue \citep{2018ApJ...859...11S}, which has more accurate SFRs from joint UV+optical+MIR SED fitting, with the medium depth of the UV photometry, hence the GSWLC-M2 catalogue. We found SFRs for 470 galaxies in the SIT (including SIT\,45, as shown in Fig.~\ref{fig:SFRmassSIT45}, where we show the values for all SIT galaxies in the GSWLC-M2 catalogue).

In Table~\ref{tab:SFRspec} we present the obtained SFR for each galaxy in SIT~45 in comparison with the values reported in the GSWLC-M2 catalogue. The values of the SFR are fairly similar for galaxy SIT~45C, but there is some discrepancy for galaxies SIT~45A and SIT~45B, as it can be appreciated in the SFR-M$_\star$ diagram in Fig.~\ref{fig:SFRmassSIT45}. These differences may be due in part because of the different input photometry and the different SED fitting priors (for instance different IMF), but mainly due to the different methodology. The GSWLC-M2 catalogue considers a two-exponentially decreasing function to take into account the contribution of a late star-formation burst event due the interaction of the member galaxies. In addition, given that galaxy SIT~45A looks overlapping SIT~45B from our point of view, it is difficult to separate the photometry contribution of each galaxy to analyse them. For this reason we use the resulted SFR map to identify the contribution of each galaxy (orange contours in Fig.~\ref{fig:sfr_map}), instead of using photometry (cyan and red contours in Fig.~\ref{fig:sfr_map}). 

Note also that the stellar mass for galaxy SIT~45B from UV-to-NIR SED fitting (log(M$_\star$)~=~10.29\,M$_\odot$) is larger than the stellar mass provided in the GSWLC-M2 catalogue, which is based on photometry for all the galaxies in the SIT (see the Fig.~\ref{fig:SFRmassSIT45}).

\begin{table*}
\centering
\begin{tabular}{ccccccc}
\hline\hline
(1) & (2) & (3) & (4) & (5) & (6) & (7)\\
Galaxy & $\tau_{main}$ & $Age_{bq}$ & $r_{SFR}$ & SFR & SFR$_{100}$ & log($M_\star$) \\ 
 & (Myr) & (Myr) & & ($M_{\odot}\rm yr^{-1}$) & ($M_{\odot}\rm yr^{-1}$) &  ($M_{\odot}$)  \\
\hline
SIT~45A & 7063\,$\pm$\,1447 & 116\,$\pm$\,102 & 1.7\,$\pm$\,0.7 & 0.51\,$\pm$\,0.10 & 0.50\,$\pm$\,0.11 & 9.70\,$\pm$\,0.04\\ 
SIT~45B & 5302\,$\pm$\,1719 & 217\,$\pm$\,108 & 0.3\,$\pm$\,0.4 & 0.31\,$\pm$\,0.15 & 0.45\,$\pm$\,0.24 & 10.29\,$\pm$\,0.05\\ 
SIT~45C & 6826\,$\pm$\,1776 & 213\,$\pm$\,84 & 5.3\,$\pm$\,1.9 & 2.62\,$\pm$\,0.38 & 2.30\,$\pm$\,0.44 & 9.84\,$\pm$\,0.08 \\
\hline
\end{tabular}
\caption{Parameters found for the SFH of SIT~45 following a Bayesian analysis as provided by CIGALE using photometry from the UV-to-NIR. The columns correspond to: (1) galaxy in the triplet SIT~45; (2) $e$-folding time of the main stellar population model, in Myr; (3) age of the burst/quench, in Myr; (4) ratio of the SFR after/before $Age_{bq}$; (5) star formation rate, in $M_{\odot}\rm yr^{-1}$ ; (6) average SFR over 100\,Myr, in $M_{\odot}\rm yr^{-1}$; (7) stellar mass, in $M_{\odot}$. Intensive parameters of the SFH ($\tau_{main}$, $Age_{bq}$, and $r_{SFR}$) are luminosity weighted.}
\label{tab:SIT45_bayesian}
\end{table*}

\begin{table*}
\centering
\begin{tabular}{cccc}
\hline \hline
(1)  & (2) & (3) & (4) \\
Galaxy  & log(SFR)$_{\rm GSWLC}$ & log(SFR)$_{\rm UV-to-NIR}$ & log(SFR)$_{\rm fiber}$\\ 
 & $M_{\odot}\rm yr^{-1}$ & $M_{\odot}\rm yr^{-1}$ & $M_{\odot}\rm yr^{-1}$ \\ 
\hline
SIT~45A & 0.277\,$\pm$\,0.077 & $-$0.290\,$\pm$\,0.083 & $-$0.534\,$\pm$\,0.148 \\ 
SIT~45B & 0.283\,$\pm$\,0.079 & $-$0.516\,$\pm$\,0.206 & $-$0.155\,$\pm$\,0.127 \\ 
SIT~45C & 0.391\,$\pm$\,0.035 & 0.418\,$\pm$\,0.064 & 0.442\,$\pm$\,0.068 \\ 
\hline
\end{tabular}
\caption{SFR for the galaxies in SIT~45. The columns correspond to: (1) Galaxy name; (2) SFR in the GSWLC-M2 catalogue \citep{2018ApJ...859...11S}; (3) SFR in this work using photometry from UV to NIR; (4) SFR in this work combining photometry (fiber magnitude) and optical SDSS spectra.}
\label{tab:SFRspec}
\end{table*}

\subsection{Environment of SIT~45}
\label{Sec:resenv}

We used stellar masses obtained from UV-to-NIR SED fitting and the values of the apparent diameters in Table~\ref{tab:SIT45_general} to estimate the $Q_{\rm trip}$ parameter for SIT~45 following Eq.~\ref{Eq:Qtrip}. We found a value of $Q_{\rm trip}=~-0.09~\pm~0.09$, which is one of the highest values for the SIT, with mean value $Q_{\rm trip}=~-2.14$ and standard deviation 0.83. This result is expected given the fact that SIT~45 is a merging system. 

According to the quantification of the LSS environment for isolated triplets in \citet{2015A&A...578A.110A}, SIT~45 presents a higher value of the tidal parameter due to the LSS environment ($Q_{\rm LSS}=-3.31$) than the rest of the SIT (with a mean value of $Q_{\rm LSS}~=~-5.05$ and standard deviation 0.71). This is mainly due to SIT~45 having 28 LSS associated galaxies, meanwhile the mean value in the SIT is 25. However the nearest neighbour is found at a projected distance $d_{nn}$~=~2.26\,Mpc (where the mean value for the SIT is $d_{nn}$~=~1.42\,Mpc). The value of its projected density ($\eta_{\rm k, LSS}=-1.20$) is comparable with the rest of the SIT with mean value $\eta_{\rm k, LSS}=-1.24$ and standard deviation 0.47. This suggests that even if SIT~45 is an isolated triplet, it might be influenced by the LSS. We have checked with the LSSGalPY \citep{2015ascl.soft05012A,2017PASP..129e8005A} tool that there is a small cluster (i.e. structure composed of $\sim$\,100 galaxies at the intersection of two filaments) at a projected distance d~$\sim$~2.49\,Mpc from SIT~45. This means that SIT~45 might be accreting of cold gas from the LSS, that could trigger nuclear activity or star formation in the centre of the galaxies, depending on their stellar mass \citep{2013MNRAS.430..638S,2016A&A...592A..30A,2018A&A...620A.113A}.

The value of $Q_{\rm trip}$ for SIT~45 is about five orders of magnitude higher than $Q_{\rm LSS}$, as expected since the galaxies are in interaction. This means that the effect of the LSS environment is negligible in comparison to the tidal strength due to the triplet member galaxies. This value is also higher than for most of the SIT triplets, also a consequence of the interaction of the galaxies and the fact that mergers are relatively unusual in isolated triplets (nine triplets out of 315), as previously introduced in Sec.~\ref{Sec:intro}.

Note that the size on SIT~45C is smaller than expected from Fig.~\ref{fig:SIT45_SDSS_grid}, as well as SIT~45A has a size that is comparable to that SIT~45B. SIT~45C is a galaxy with a bright stellar nucleus, therefore its measured Petrosian flux misses most of the extended light of the object and its Petrosian radius is set by its nucleus alone \citep{2001AJ....121.2358B,2001AJ....122.1104Y}. 
In the case of SIT~45A, its apparent size is overestimated due to the difficulty of separating the photometry for overlapping galaxies. According to Eq.~\ref{Eq:Q}, the smaller the apparent diameter the higher $Q$ parameter, therefore the values of $Q_{\rm trip}$ and $Q_{\rm LSS}$ are underestimated by $\sim$\,0.2\,dex. This discrepancy does not have any effect on our results and conclusions, since we have checked that the nine SIT triplets with ongoing mergers have $Q_{\rm trip}$\,>\,$-$0.5 (V\'asquez-Bustos et al. in preparation), with SIT~45 one of the four triplets with the highest $Q_{\rm trip}$, as shown in Fig.~\ref{fig:QtripDynParam}.

\subsection{Dynamics and configuration of SIT~45}
\label{Sec:resdyn}

\begin{figure}
\centering
\includegraphics[width=\columnwidth]{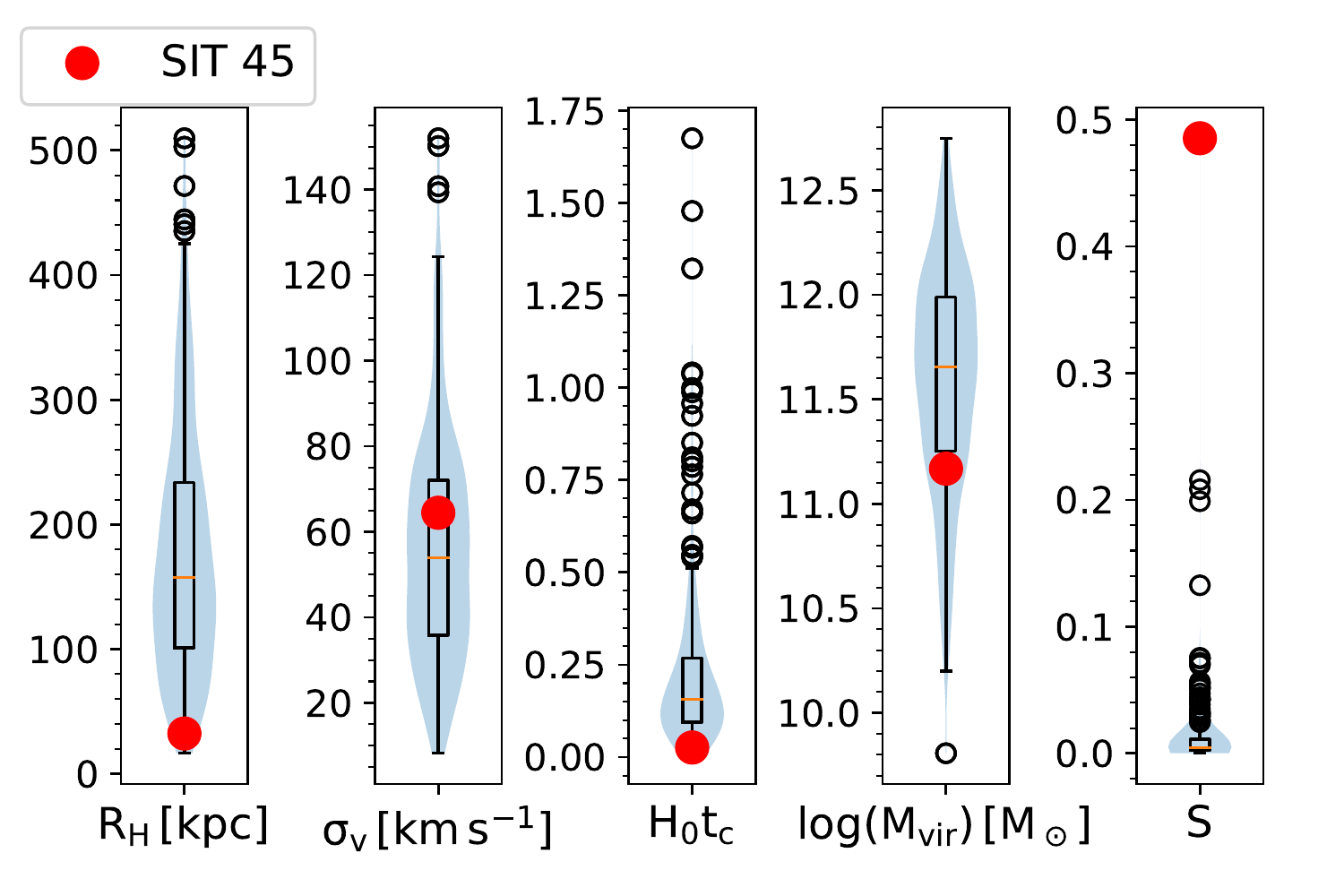}
\caption{Comparison of the dynamical parameters found for SIT~45 (red filled circles) with respect to the SIT. From left to right the projected harmonic radius ($R_H$), velocity dispersion ($\sigma_{v}$), crossing time ($H_{0}t_{c}$), virial mass ($M_{\rm vir}$), and compactness of the triplet (S). The pale blue violin shape corresponds to a density diagram, rotated and placed on each side, to show the form of data distribution for each dynamical parameter. The box inside the main body of the violin diagram shows the interquartile range (IQR) of the median (represented by the orange horizontal line) and its 95\% confidence intervals. The vertical lines starting from the inner box correspond to 1.5~$\times$~IQR. The atypical values of the distribution are represented by black open circles.}
\label{fig:dynparams_SIT45}
\end{figure}

We used the dynamical parameters projected harmonic radius, $R_H$; velocity dispersion, $\sigma_{v}$; crossing time, $H_{0}t_{c}$; and virial mass of the triplet, $M_{\rm vir}$, described in Sec.~\ref{Sec:dynparams} to study the dynamics of SIT~45 in a cosmological scenario where the system is built in a common primordial dark matter halo. As mentioned, we also used the compactness $S$ to study the configuration of the triplet. The values of these parameters for SIT~45 are shown in Table~\ref{tab:SIT45_dinamic}. Figure~\ref{fig:dynparams_SIT45} shows the dynamical parameters of SIT~45 in comparison with the distributions for the rest of the triplets in the SIT using statistical violin plots. 

As can be seen in Fig.~\ref{fig:dynparams_SIT45}, the value of $\sigma_{v}$ for SIT~45 ($\sigma_{v}$~=~64.43\,$\pm$\,0.65\,km\,s$^{-1}$) is slightly higher than the median value in the SIT ($\sigma_{v}$~=~53.9\,km\,s$^{-1}$) but inside the interquartile range, which indicates that SIT~45 is in agreement with the distribution of $\sigma_{v}$ for isolated triplets, in general, within 40\,km\,s$^{-1}$ and 60\,km\,s$^{-1}$ approximately. On the contrary, the values of $R_H$, $H_0t_c$, and $\rm M_{\rm vir}$ for SIT~45 ($R_H$~=~32.38\,$\pm$\,0.03\,kpc, $H_0t_c$~=~0.0259\,$\pm$\,0.0003, and $\rm log(M_{vir})$~=~11.17\,$\pm$\,0.87\,M$_\odot$) are lower than the typical values in the SIT (with median values $R_H$~=~157.6\,kpc, $H_0t_c$~=~0.156, and $\rm{log} (M_{vir})$~=~11.65\,M$_\odot$), and even lower than the minimum value of their interquartile range. Regarding the parameters $R_H$ and $H_0t_c$, the results suggest that SIT~45 has a more compact structure than the rest of the isolated triplets. Considering its virial mass, SIT~45 is one of the least massive triplets in the SIT, with a percentage of dark matter\footnote{The percentage of Dark Matter (DM) content is estimated using $M_{vir} = M_{\rm \star, total} + M_{\rm DM}$, with $M_{\rm \star, total}~=~M_{\rm \star, SIT45A} + M_{\rm \star, SIT45B} + M_{\rm \star, SIT45C}$, using stellar masses provided in Table~\ref{tab:SIT45_bayesian} estimated from SED fitting (Sec.~\ref{Sec:SED}).} of about 78.6\%, which is in agreement with the standard cosmological model.

We also use the $S$ parameter, defined in Eq.~\ref{eq:SS} following \citep{2013MNRAS.433.3547D}, to consider the compactness of the triplet. We estimated $S$ by defining a circle, $r_{m}$, encompassing the centre of the three galaxies, finding a value of with $r_{m}$~=~55\farcs45 for SIT~45. The compactness of SIT~45 is $S$\,=\,0.48\,$\pm$\,0.17, being the maximum value for the SIT, with median $S$\,=\,0.0045 and standard deviation $S$\,=\,0.0364.

\begin{figure}
\centering
\includegraphics[width=\columnwidth,trim={0cm 0cm 1cm 1cm},clip]{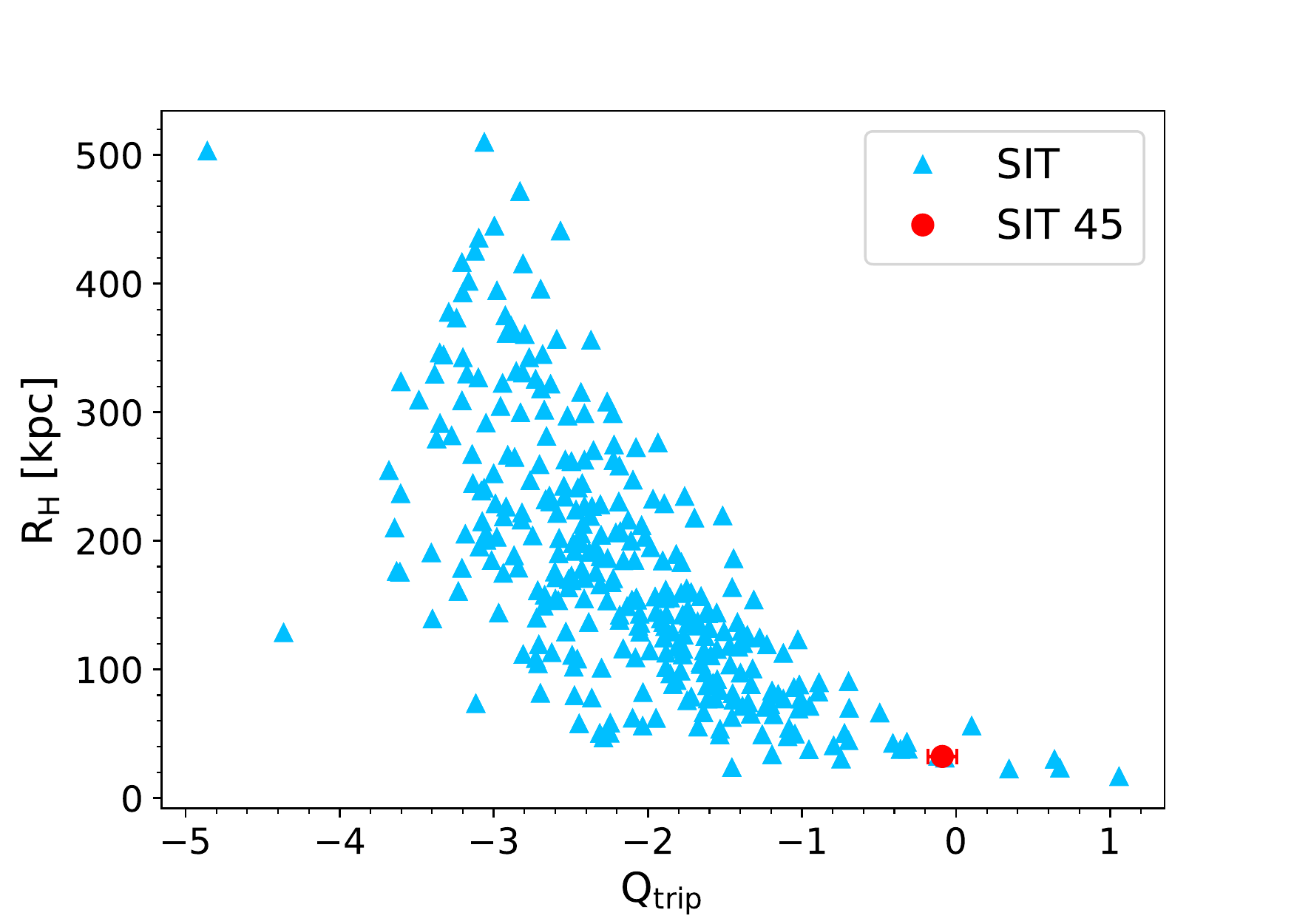}
\includegraphics[width=\columnwidth,trim={0cm 0cm 1cm 1cm},clip]{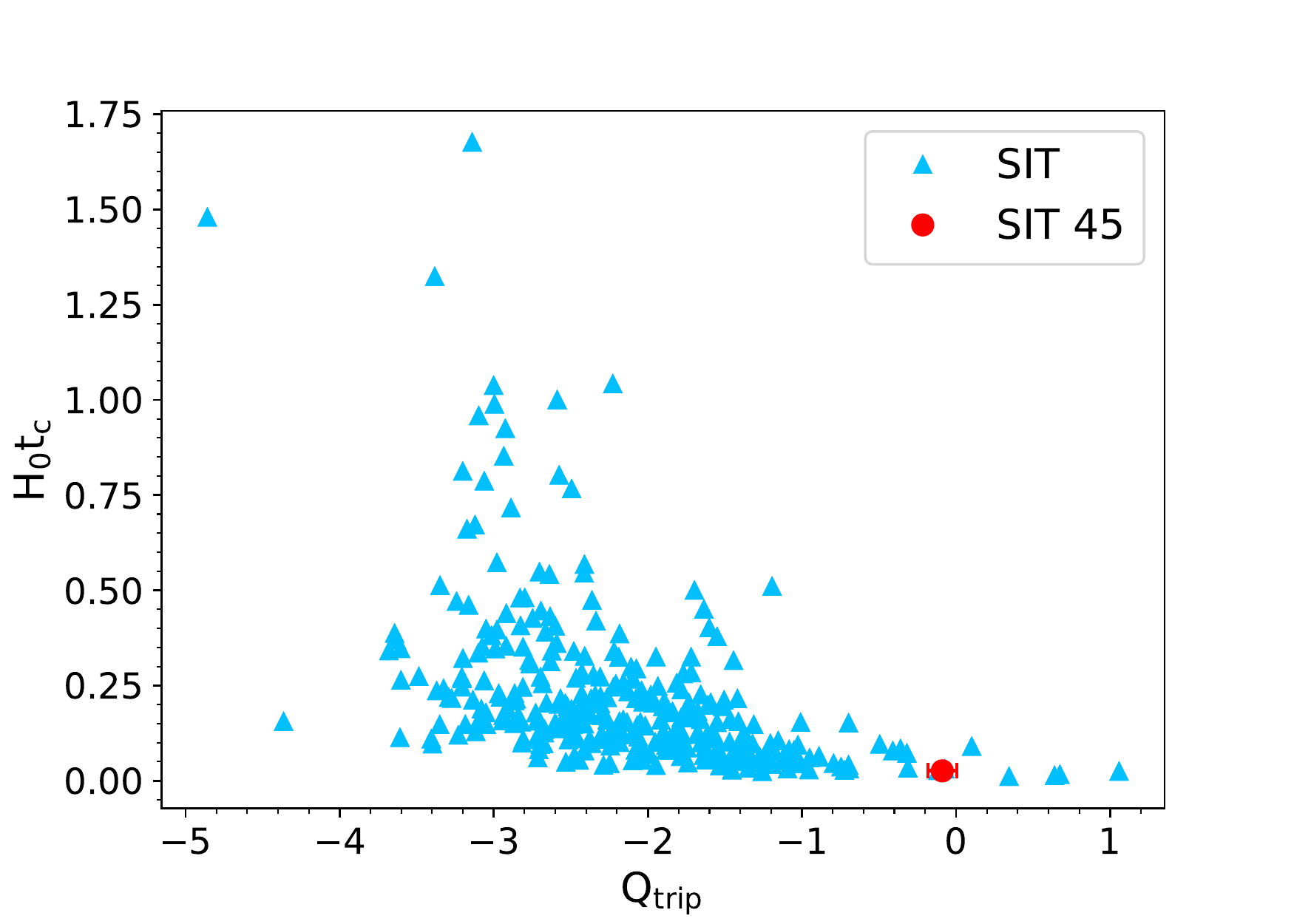}
\caption{Relation found of the dynamical parameters $R_H$ (upper panel) and $H_0t_c$ (lower panel) with the tidal strength exerted by the triplet member galaxies, $Q_{trip}$. The values represented by light blue triangles correspond to the values for the triplets in the SIT. The values corresponding to the isolated triplet SIT~45 are depicted by a red circle, with their corresponding errors using red error bars.}
\label{fig:QtripDynParam}
\end{figure}

SIT~45 is one of the triplets with higher $Q_{\rm trip}$ and lower $R_H$ and $H_0t_c$, as shown in Fig.~\ref{fig:QtripDynParam}. According to V\'asquez-Bustos et al. (in prep.), it means that SIT~45 is a strongly interacting triplet but it does not necessarily correspond to a compact and long-term evolved system. We will discuss these results further in Sec.~\ref{sec:disdyn}.

\begin{table*}
    \centering
    \begin{tabular}{ccccc}
    \hline\hline
    (1) & (2) & (3) & (4) & (5)\\
     $R_{\rm H}$ & $\sigma_v$  & $H_0t_c$  & $\rm log(M_{\rm vir})$ & $S$ \\ 
     (kpc) & (km\,s$^{-1}$)  &  & ($M_{\odot}$) \\ 
     \hline
     32.38\,$\pm$\,0.03 & 64.43\,$\pm$\,0.65 & 0.0259\,$\pm$\,0.0003 & 11.17\,$\pm$\,0.87 & 0.48\,$\pm$\,0.17\\ 
     \hline
     \end{tabular}
    \caption{Dynamical parameters of SIT~45. The columns correspond to: (1) harmonic radius; (2) velocity dispersion; (3) crossing time; (4) virial mass; (5) compactness.}
    \label{tab:SIT45_dinamic}
\end{table*}

\section{Discussion}
\label{Sec:dis}

\subsection{Star formation history of SIT~45} 
\label{sec:dis_sfh}

To try to infer the dynamical history of SIT~45 with its assembly history, we derived physical properties as stellar age, SFR, and dust attenuation of the stellar populations for each galaxy of the triplet, using state-of-the-art UV/optical/IR SED fitting techniques. We considered photometry from GALEX, SDSS, 2MASS, and unWISE. As indicated in Sec.~\ref{Sec:SED}, we constrained the SFH assuming a delayed SFH with optional constant burst/quench event due the interaction of the member galaxies.

The resulting physical parameters of the modelled SFH with CIGALE are shown in Table~\ref{tab:SIT45_bayesian}. The second column corresponds to the obtained values of the $\tau_{main}$ parameter of the SFH. It shows that the $e$-folding time of the main stellar population in SIT~45B is smaller than in SIT~45A and SIT~45C, being SIT~45C the galaxy with the highest value. The values of $r_{SFR}$ indicate recent star formation in the system. There is a burst for SIT~45B and SIT~45C $\sim$220\,Myr ago, with a significant increase of the SFR in SIT~45C with respect to SIT~45B, which is much more modest. This event might refer to a previous encounter, leading to the tidal morphologies observed in the system. There is a recent burst episode in SIT~45A, which may have happened about 100\,Myr after the encounter between SIT~45B and SIT~45C, however the uncertainty is large and it might have happened even at the same time as for galaxies SIT~45B and SIT~45C. 

According to the values of the SFR, the three galaxies have ongoing star formation, with SIT~45C being the galaxy with the highest SFR. In fact, the starburst scenario for SIT~45C is supported when taking into account the stellar mass of the galaxy. In comparison with the rest of the triplets in the SIT, the SFR-M$_\star$ diagram (presented in Fig.~\ref{fig:SFRmassSIT45}) shows that SIT~45C is located above the envelope of the star-formation main sequence, while SIT~45A and SIT~45B are located in the main sequence. These recent star-forming, with the starburst scenario for SIT~45C, are likely connected with the merging process in the system. 

As expected, the uncertainties are larger in the ratio of the SFR after/before the age of the burst/quench in young burst and its age, but not so much in the stellar mass and SFR. 
In order to get more accurate physical parameters it is necessary to take into account other parameters to constrain the SFH, such as the $H_\alpha$/$H_\beta$ ratio or higher resolution IR photometry to better constraint the dust attenuation and minimise its effect in the SED modelling. While we do not have higher resolution IR photometry for the system, there are public optical SDSS spectra of the inner (3.5\arcsec) region of the galaxies (see Fig.~\ref{fig:SIT45spectra}). We have compared the obtained SFR with the SFR after considering the SDSS spectrum of each galaxy, in combination with photometry (fiber magnitudes). We also used CIGALE, which not only fits passband fluxes but also physical properties and can also fit emission lines fluxes. In this case we used the D4000 parameter and the $H_\alpha$ and $H_\beta$ emission lines fluxes. We found that the resulting SFR is consistent, as shown in Table~\ref{tab:SFRspec} and Fig.\ref{fig:SFRmassSIT45}. The results are relatively consistent to the expected values of the SFR at fixed stellar mass according to the D$_n$(4000)\footnote{Narrow definition of the 4000\,$\AA$ break, as defined in \citet{1999ApJ...527...54B}, as a proxy of the main age of the stellar population.} parameter of each galaxy as presented in \citet{2022arXiv220501203D}. The SFR for galaxies SIT~45B and SIT~45C are within the expected range according to their D$_n$(4000) values  \citep[D$_n$(4000)~=~1.38, and D$_n$(4000)~=~1.26, respectively, from the measurements performed in][]{2004MNRAS.351.1151B}. However, as shown in Fig.\,7 in \citet{2022arXiv220501203D}, galaxies with younger stellar populations (Dn(4000) < 1.2) present higher SFR. Therefore, the expected SFR for galaxy SIT~45A (D$_n$(4000)~=~1.06) should be higher than the value we have found. Overall, our results show that SIT~45 presents high values of SFR for stellar masses in the range of 8\,<\,log(M$_\star$)\,<\,11.5\,M$_\odot$ (as shown in Fig.~\ref{fig:SFRmassSIT45}), being therefore classified as star-forming galaxies.

According to the morphology of the galaxies in SIT~45 (blue spiral galaxies), and supported by their spectroscopic properties (star-forming galaxies), SIT~45 is therefore a major ``wet'' merger, i.e. a merger of similar stellar mass rich-gas blue galaxies. Our results are in agreement with \citet{2013MNRAS.433.3547D}, where a sample of 71 triplets from the SDSS was studied. They found that blue triplets (which in general are less massive) show efficient total star-formation activity in comparison to control samples.

In general, wet mergers have been found to be more prominent in low-density environments \citep{2001PASJ...53..395B, 10.1111/j.1365-2966.2009.15557.x,Lin_2010}. It is important to note that we have visually inspected all the SIT triplets and we have found only 9 ``wet'' mergers (including SIT~45). SIT~45 is an excellent example of a triplet that might have been caught in the act of merging, before stripping and/or gas consumed to become earlier type galaxies, i.e. not so evolved. The existence of this type of systems can be considered as a consequence of the formation of structures, which occurs more slowly and on smaller scales than in regions with medium density, and a possible way for the formation of gas-rich discs \citep{10.1093/mnras/stw2841}. It is therefore an ideal candidate to study short time-scale mechanisms ($\sim 10^8$\,years), such as interaction triggered star-formation or starbursts induced by galaxy--galaxy interactions which are more frequent at higher redshift. 

We also considered the possibility of an additional scenario where SIT~45A is a tidal galaxy formed from the interaction between SIT~45B and SIT~45C.
According to its SFH, there is almost no old stellar population in SIT~45A, with no detection, or very faint, in 2MASS and WISE photometry (see Fig.~\ref{fig:multilambda}). On the contrary, the galaxy is very bright in FUV and NUV, indicating a recent ($\sim$\,100\,Myr) increase of its star formation, and in line with its SFH, SIT~45A would be still forming new stars, supporting the tidal galaxy scenario. To further explore this scenario it would be necessary to perform a better SED fitting considering physical properties and emission lines, as well as an analysis of the gas-phase metallicities in extended regions, including tidal tails and bridges between galaxies SIT~45B and SIT~45C. We discarded the possibility of SIT~45A being an HII region in the tidal arm of SIT~45B. Our assumption is supported by the fact that their redshift difference even larger than the difference between SIT~45A and SIT~45C, therefore SIT~45A is not bound to SIT~45B, contrary to what is observed in HII regions \citep{2013A&A...552A.140R}. Additionally, the images of the galaxy with deeper, higher resolution photometry (as Subaru HSC), shows SIT~45A as an individual galaxy, with observable different stellar populations (colours), rather than a structure of SIT~45B. This hypothesis would be easily confirmed from the kinematic analysis of SIT~45A with respect to SIT~45C, ideally using optical deep integral field spectroscopy to be able to also better constraint their SFH.

\subsection{Dynamics and configuration of SIT~45} 
\label{sec:disdyn}

Given that SIT~45 is a merging system, we expect it to show complex dynamics. To quantify and understand its dynamical evolution we estimated its dynamical parameters $R_H$, $\sigma_{v}$, $H_0t_c$, and $M_{vir}$, described in Sec.~\ref{Sec:dynparams}, and presented in Sec.~\ref{Sec:resdyn}.

With a value of $R_H$~=~32.2\,kpc, SIT~45 is one of the most compact triplets in the SIT, with a median value of $R_H$~=~157.6\,kpc (as shown in Fig.~\ref{fig:dynparams_SIT45}). It is also more compact than triplets in other samples, with median values between $R_H$~=~72\,kpc and $R_H$~=~191\,kpc \citep{2000ARep...44..501K,2000ASPC..209...40M,2005KFNT...21a...3V,2009MNRAS.394.1409E}. Note that median values found for triplets are also comparable with values found in compact groups \citep{1992ApJ...399..353H,2015MNRAS.447.1399D}. SIT~45 has a velocity dispersion $\sigma_{v}$~=~64.5\,km\,s$^{-1}$, which is comparable with the median value of the SIT ($\sigma_{v}$~=~53.9\,km\,s$^{-1}$). However, SIT triplets have lower velocity dispersion than other galaxy triplets and compact groups, which show velocity dispersion values $\sigma_{v}$~$\sim$~200\,km\,s$^{-1}$, \citep{2015MNRAS.447.1399D}, but the values are comparable with galaxy pairs and triplets selected by a dynamical methods \citep{2000ASPC..209...40M,2005KFNT...21a...3V,2009MNRAS.394.1409E}. 

We estimated a complementary configuration parameter, the compactness $S$, as defined in \citet{2013MNRAS.433.3547D}. They calculated $S$ for a sample of galaxy triplets selected from the SDSS, and related it with their total stellar mass, obtaining values in the range 0.03~$\lesssim$~$S$~$\lesssim$~0.35. The higher the value of $S$, the more compact the triplet (see Eq.~\ref{eq:SS}). This parameter ranges from $S$\,=\,0.0001 to $S$\,=\,0.48 for the SIT, being SIT~45 the most compact isolated triplet. The estimated compactness for SIT~45 is therefore in agreement with the suggestion that it is a very compact triplet from $R_H$, and it seems to be even more compact than the triplets considered in \citet{2013MNRAS.433.3547D}. However, they observed that $S$ increases with the total stellar mass of the system, so that the blue (and therefore less massive) triplets tend to present less compact configurations. For instance, the compact isolated triplet J0848\,+\,1644 composed of three early-type galaxies and with $R_H$\,=\,14.6\,kpc \citep{2016RAA....16...72F} satisfies this trend. In the case of SIT~45 (composed of three low-mass blue galaxies) its high compactness might be due to its merging stage.

According to \citet{1992ApJ...399..353H}, the dimensionless crossing time $H_0t_c$ is a convenient measurement of the maximum of the number of times a galaxy could have traversed the group since its formation. It is therefore an estimation of the dynamical state of a system, where systems with a long-term evolution show low values of $H_0t_c$. \citet{1992ApJ...399..353H} found a significant correlation between crossing time (with a median value of 0.016$H_0^{-1}$) and the fraction of gas-rich galaxies in compact groups, where groups with low values of $H_0t_c$ typically contain fewer late-type galaxies. This trend might suggest that groups with low $H_0t_c$ would be dynamically more evolved. This hypothesis agrees, for instance, with the study done by \citet{2016RAA....16...72F} on the isolated compact galaxy triplet J0848\,+\,1644, composed of three early-type galaxies, with $H_0t_c\,=\, 0.032$. \citet{2015MNRAS.447.1399D} also found similar results, both from a catalogue of galaxy triplets and simulations, where the formation of early-type galaxies in evolved systems may have been favoured by galaxy mergers, in agreement with \citet{1989Natur.338..123B} for compact groups. 

Given the nature of SIT~45, composed of three blue late-type (gas rich) spirals, would not be considered as a highly evolved system, however the value of its dimensionless crossing time is one of the lowest in the SIT ($H_0t_c\,=\,0.026$, see Table~\ref{tab:SIT45_dinamic}). This result would imply that SIT~45 is generally more dynamically evolved compared to the SIT. However this result might be affected by the dependence of the $H_0t_c$ estimation with $R_H$, where SIT~45 presents also a very low value with respect to the SIT, and by consequence it will show a low value of $H_0t_c$. In this sense, we also have found that these two values are somehow connected with the local environment. 

In general, we have not observed a clear dependence of the dynamical parameters with the LSS environment. This result is in agreement with \citet{2015A&A...578A.110A}, who found that the LSS environment is less relevant than the local environment, where local neighbours typically exert about the 99\% of the total tidal strength. However we do find a clear relation for $R_H$ and $H_0t_c$ with the local environment, quantified by the $Q_{\rm trip}$ parameter (as shown in the upper and lower panels of Fig.~\ref{fig:QtripDynParam} for $R_H$ and $H_0t_c$, respectively). High values of $R_H$ and $H_0t_c$ are restricted to systems with low values of $Q_{\rm trip}$, i.e. no major tidal forces exist between triplet galaxies. In addition, isolated SIT triplets with lower values of the dynamical parameters span a wider range of $Q_{\rm trip}$ values, where all the systems with higher $Q_{\rm trip}$ (with values above $-0.5$, more likely strongly interacting systems) are in the range of the lowest $R_H$ and $H_0t_c$ (V\'asquez-Bustos et al. in preparation). These results contradict the findings in \citet{1994ApJ...427..684M} for compact groups, where no correlation was found between the number of interacting galaxies and the crossing time (or group velocity dispersion).

We conclude that more dynamically evolved systems show low $H_0t_c$, but it does not necessary imply a relation in the opposite direction, it depends on the degree of the influence of the member galaxies. Therefore the observed results for SIT~45 would be connected with its ongoing merging stage rather than its dynamical evolution stage, in agreement with V\'asquez-Bustos et al. in preparation.  

The conclusion is also supported by the dynamical mass of SIT~45, with a value of $\rm log(M_{vir})\,=\,10.6\,M_\odot$ it is one of the least massive SIT triplets. Using the stellar mass of each galaxy in SIT~45 we have estimated a 78.6\% of dark matter in the system. This value is in agreement with the standard cosmological model, which suggests that the three member galaxies might be embedded in a common dark matter halo \citep{1992SvA....36..231A,2005AN....326..502K}, favouring the dynamical evolution of SIT~45. The mock analysis for galaxy triplets in \citet{2015MNRAS.447.1399D} also support this assumption, where all the identified triplets have galaxy members that belong to the same dark matter halo, which is a strong evidence of the dynamical co-evolution of the system.


\section{Summary and conclusions}
\label{Sec:con}

In this work we have studied the dynamical parameters and SFH of the isolated merging galaxy triplet SIT~45. The interacting system SIT~45 (UGC~12589) is an unusual isolated triplet of galaxies, consisting of three interacting late-type galaxies. It is therefore an ideal candidate for investigating processes such as the triggering of star formation due to interaction.

To study the dynamical evolution of SIT~45 we used dynamical parameters harmonic radius $R_H$, velocity dispersion $\sigma_{v_r}$, crossing time $H_0t_c$, and virial mass $\rm log(M_{vir})$. We have explored the connection of its dynamical evolution with its local and large-scale environments, characterised by the tidal strength parameter $Q$ and the projected local density $\eta_k$.

To relate the dynamical history of SIT~45 with its assembly history, we derived stellar age, SFR and dust attenuation of the stellar populations to constrain the SFH for each galaxy of SIT~45, using state-of-the-art UV/optical/IR SED fitting techniques.

Our main conclusions are the following:

\begin{enumerate}
\item On one hand, the SIT~45 triplet is a highly isolated system with respect to its large-scale environment. On the other hand, the value of its tidal force parameter due to triplet members is one of the highest in the SIT. The system is compact as shown by its harmonic radius and compactness, which is consistent with the fact that the three galaxies that compose it are  interacting.

\item According to its star formation history, the system has ongoing star-formation, with SIT~45C presenting \textit{starburst} activity. The galaxies present recent ($\sim $200\,Myr) star formation increase, indicating that it may have been triggered by the ongoing merging process. 

\item The harmonic radius and crossing time values are much smaller than in the rest of the SIT triplets, which would suggest that the system is highly evolved. However, contrary to what would be expected, the triplet is composed of blue spirals with high star formation rate.

\item The percentage of dark matter of the triplet, estimated from its virial and stellar mass, suggests that the three member galaxies might be embedded in a common dark matter halo.

\item Taking into account these results, together with the fact that its velocity dispersion has a value similar to that of the SIT triplets, we conclude that SIT~45 is a system of three interacting galaxies that are evolving within a common dark matter halo. The isolated triplet SIT~45 is therefore an ideal candidate to study short time-scale mechanisms ($\sim 10^8$ years), such as interaction triggered star-formation or starbursts induced by galaxy--galaxy interactions which should be more frequent at higher redshift.
\end{enumerate}

Considering our results, we propose two different scenarios for the formation of SIT~45A. 

The analysis of the SFH from multi-wavelength SED fitting suggests the scenario where SIT~45A might be a tidal galaxy formed from the interaction between SIT~45B and~SIT 45C. Tidal dwarf galaxies (TDGs) are formed of torn off material from the outer parts of a spiral disk due to tidal forces in a collision between two massive galaxies \citep{2001A&A...378...51B,2003A&A...402..921D}. According to numerical simulations, 25\% become long-lived bound objects that typically survive more than 2\,Gyr with masses above $10^8\,M_\sun$ \citep{2006A&A...456..481B}. Other works suggest that TDGs have stellar masses in the range of $7.5\,\leq\,\log(M_\star/M_\sun)\leq9.5$ \citep{2004A&A...427..803D,2020MNRAS.499.3399R}. SIT~45A has an estimated mass of $\log(M_\star/M_\sun)\,=\,9.70\,\pm\,0.04$, therefore is slightly above this range and would not be considered as a dwarf galaxy. However, this estimation might be overestimated given the difficulty of separating the stellar populations of SIT~45A, which overlaps with SIT~45B with respect to our line-of-sight.

Nevertheless, we also propose another scenario where the interaction between SIT~45B and SIT~45C occurs before the arrival of SIT~45A. This is consistent with the redshift differences between galaxies in the system, where SIT~45A has higher radial velocity difference with respect to SIT~45B and SIT~45C. In merging pairs, there is an excess of young stellar population directly related to the ongoing merging process \citep{2012A&A...539A..45L}. However, on the other hand, galaxies in pairs with tidal features show evidence of older stellar populations that can be associated to a larger time-scale of the interaction. In this scenario, \citet{2018MNRAS.481.2458D} found enhanced star formation indicators for  galaxies in triplets that have a close companion. Therefore, the recent arrival of SIT~45A might also contribute to the increment of star formation activity, supporting this scenario. 

To further explore the proposed scenarios and the kinematic mechanisms that triggered star formation in SIT~45, integral field spectroscopy observations would be necessary to allow investigating the SFR, SFH, and stellar velocity and velocity dispersion in extended regions for each galaxy. Since TDGs are found to form in massive gaseous accumulations, CO and HI observations would be necessary to study the gas component of the system. Our results point toward the necessity of developing a better understanding of the dynamics of galaxy triplets.

\begin{acknowledgements}
We thank our referee whose valuable comments have certainly contributed to improve and clarify this paper.

MAF and PVB acknowledge financial support by the DI-PUCV research project 039.481/2020. MAF also acknowledges support from FONDECYT iniciaci\'on project 11200107 and the Emergia program (EMERGIA20\_38888) from Consejería de Transformación Económica, Industria, Conocimiento y Universidades and University of Granada. 
UL and DE acknowledge support from project PID2020-114414GB-100, financed by MCIN/AEI/10.13039/501100011033. DE also acknowledges support from Beatriz Galindo senior fellowship (BG20/00224) financed by the Spanish Ministry of Science and Innovation, and project PID2020-114414GB-100 financed by MCIN/AEI/10.13039/501100011033. 
UL, SV and DE acknowledge support from project P20\_00334 financed by the Junta de Andaluc\'ia and from FEDER/Junta de Andaluc\'ia-Consejer\'ia de Transformaci\'on Econ\'omica, Industria, Conocimiento y Universidades/Proyecto A-FQM-510-UGR20. 
MB gratefully acknowledges support by the ANID BASAL project FB210003 and from the FONDECYT regular grant 1211000. 
SDP is grateful to the Fonds de Recherche du Québec - Nature et Technologies and acknowledges financial support from the Spanish Ministerio de Econom\'ia y Competitividad under grants AYA2016-79724-C4-4-P and PID2019-107408GB-C44, from Junta de Andaluc\'ia Excellence Project P18-FR-2664, and also acknowledges support from the State Agency for Research of the Spanish MCIU through the `Center of Excellence Severo Ochoa' award for the Instituto de Astrof\'isica de Andaluc\'ia (SEV-2017-0709).

This research made use of \textsc{astropy}, a community-developed core \textsc{python} ({\tt http://www.python.org}) package for Astronomy \citep{2013A&A...558A..33A}; \textsc{ipython} \citep{PER-GRA:2007}; \textsc{matplotlib} \citep{Hunter:2007}; \textsc{numpy} \citep{:/content/aip/journal/cise/13/2/10.1109/MCSE.2011.37}; \textsc{scipy} \citep{citescipy}; and \textsc{topcat} \citep{2005ASPC..347...29T}. This research made use of \textsc{astrodendro}, a Python package to compute dendrograms of Astronomical data (http://www.dendrograms.org/).

This research has made use of the NASA/IPAC Extragalactic Database, operated by the Jet Propulsion Laboratory of the California Institute of Technology, un centract with the National Aeronautics and Space Administration.

Funding for SDSS-III has been provided by the Alfred P. Sloan Foundation, the Participating Institutions, the National Science Foundation, and the U.S. Department of Energy Office of Science. The SDSS-III Web site is http://www.sdss3.org/. The SDSS-IV site is http://www/sdss/org. Based on observations made with the NASA Galaxy Evolution Explorer (GALEX). GALEX is operated for NASA by the California Institute of Technology under NASA contract NAS5-98034. This publication makes use of data products from the Wide-field Infrared Survey Explorer, which is a joint project of the University of California, Los Angeles, and the Jet Propulsion Laboratory/California Institute of Technology, funded by the National Aeronautics and Space Administration.
\end{acknowledgements}

\bibliography{Bib}
\bibliographystyle{aa}

\end{document}